\documentclass[aoas,MSNbibl,nameyear,rotating,dvips]{arximspdf}
\usepackage{dcolumn}
\usepackage{graphicx}

\doi{10.1214/12-AOAS613} 
\volume{7}
\issue{2}
\pubyear{2013}
\firstpage{763}
\lastpage{798}

\makeatletter
\newcolumntype{d}[1]{D{.}{.}{#1}}

\newcommand{\boldbeta}{\bolds{\beta}}
\newcommand{\boldphi}{\bolds{\phi}}

\newcommand{\boldPhi}{\bolds{\Phi}}

\newcommand{\boldrho}{\bolds{\rho}}
\newcommand{\boldPi}{\bolds{\Pi}}
\newcommand{\boldSigma}{\bolds{\Sigma}}

\newcommand{\boldLambda}{\bolds{\Lambda}}
\newcommand{\boldupsilon}{\bolds{\upsilon}}

\newcommand{\boldalpha}{\bolds{\alpha}}
\newcommand{\bolddelta}{\bolds{\delta}}
\newcommand{\boldDelta}{\bolds{\Delta}}
\newcommand{\boldmu}{\bolds{\mu}}
\newcommand{\boldxi}{\bolds{\xi}}

\newcommand{\boldXi}{\bolds{\Xi}}

\newcommand{\boldzeta}{\bolds{\zeta}}

\newcommand{\boldeps}{\bolds{\varepsilon}}
\newcommand{\boldeta}{\bolds{\eta}}

\newcommand{\boldc}{\mathbf{c}}

\makeatother

\begin{document}
\begin{frontmatter}

\title{Modeling US housing prices by spatial dynamic structural
equation models}
\runtitle{Modeling US housing prices by SD-SEM}

\begin{aug}
\author[A]{\fnms{Pasquale} \snm{Valentini}\corref{}\ead[label=e1]{pvalent@unich.it}},
\author[A]{\fnms{Luigi} \snm{Ippoliti}\ead[label=e2]{ippoliti@unich.it}}
\and
\author[A]{\fnms{Lara} \snm{Fontanella}\ead[label=e3]{lfontan@unich.it}}
\runauthor{P. Valentini, L. Ippoliti and L. Fontanella}
\affiliation{University of Chieti-Pescara}
\address[A]{
Department of Economics\\
University of Chieti-Pescara\\
viale Pindaro, 42\\
65127 Pescara\\
Italy\\
\printead{e1}\\
\phantom{E-mail:\ }\printead*{e2}\\
\phantom{E-mail:\ }\printead*{e3}}
\end{aug}

\received{\smonth{11} \syear{2011}}
\revised{\smonth{11} \syear{2012}}

%
\begin{abstract}
This article proposes a spatial dynamic structural equation model for
the analysis of housing prices at the State level in the USA. The study
contributes to the existing literature by extending the use of dynamic
factor models to the econometric analysis of multivariate lattice data.
One of the main advantages of our model formulation is that by modeling
the spatial variation via spatially structured factor loadings, we
entertain the possibility of identifying similarity ``regions'' that
share common time series components. The factor loadings are modeled as
conditionally independent multivariate Gaussian Markov Random Fields,
while the common components are modeled by latent dynamic factors.
The general model is proposed in a state-space formulation where both
stationary and nonstationary autoregressive distributed-lag processes
for the latent factors are considered. For the latent factors which
exhibit a common trend, and hence are cointegrated, an error correction
specification of the (vector) autoregressive distributed-lag process is
proposed.
Full probabilistic inference for the model parameters is facilitated by
adapting standard Markov chain Monte Carlo (MCMC) algorithms for
dynamic linear models to our model formulation. The fit of the model is
discussed for a data set of 48 States for which we model the
relationship between housing prices and the macroeconomy, using State
level unemployment and per capita
personal income.
\end{abstract}

%
\begin{keyword}
\kwd{House prices}
\kwd{Bayesian inference}
\kwd{dynamic factor models}
\kwd{spatio-temporal models}
\kwd{cointegration}
\kwd{lattice data}
\end{keyword}

\end{frontmatter}

\section{Introduction}\label{intro}
This paper is concerned with the modeling of housing prices at the
State level in the US. Housing is a massive factor in people's
consumption. For industrialized nations, for example, it is the biggest
component in the basket of goods used for calculating the consumer
price index. Also, the Bureau of Labor Statistics has estimated in 2010
that about 24 percent of the total consumption of American home owners
goes toward housing. Hence, housing is big enough to leave a sizable
footprint on the economy in general.

In the generic sense, housing is also an important social institution
in our society. Not only does housing play a major role in any nation's
economy, but it also provides people with the social values of shelter,
security, independence, privacy and amenity. The state of the current
economy and recent events in the housing sector have thus led to
increased attention on the role of the housing sector in the economy as
a whole.

Economists have studied the relationship between the housing sector and
the macroeconomy since the 1970s. Several socio-economic variables
and/or real estate characteristics are traditionally considered to have
an impact on housing prices and several studies have thus been
dedicated to the determination of fundamental factors explaining US
housing price variations. Our primary purpose here is not to
comprehensively examine all these variables. In fact, there is no
single generally agreed upon set of variables used in testing models of
housing prices in the
literature. For a complete discussion on this point see, for example,
\citet{Mal99}, \citet{Capetal02} and \citet{Gal08}. It is thus
beyond the scope of this paper to discuss the possible roles played by
all fundamental factors in explaining the variation of housing prices.
Hence, for simplicity, we only examine here the extent to which these
prices are driven by the real per capita disposable income and the
unemployment rate.

\subsection{The data: A brief description}\label{subsecdata}

\begin{sidewaysfigure}

\includegraphics{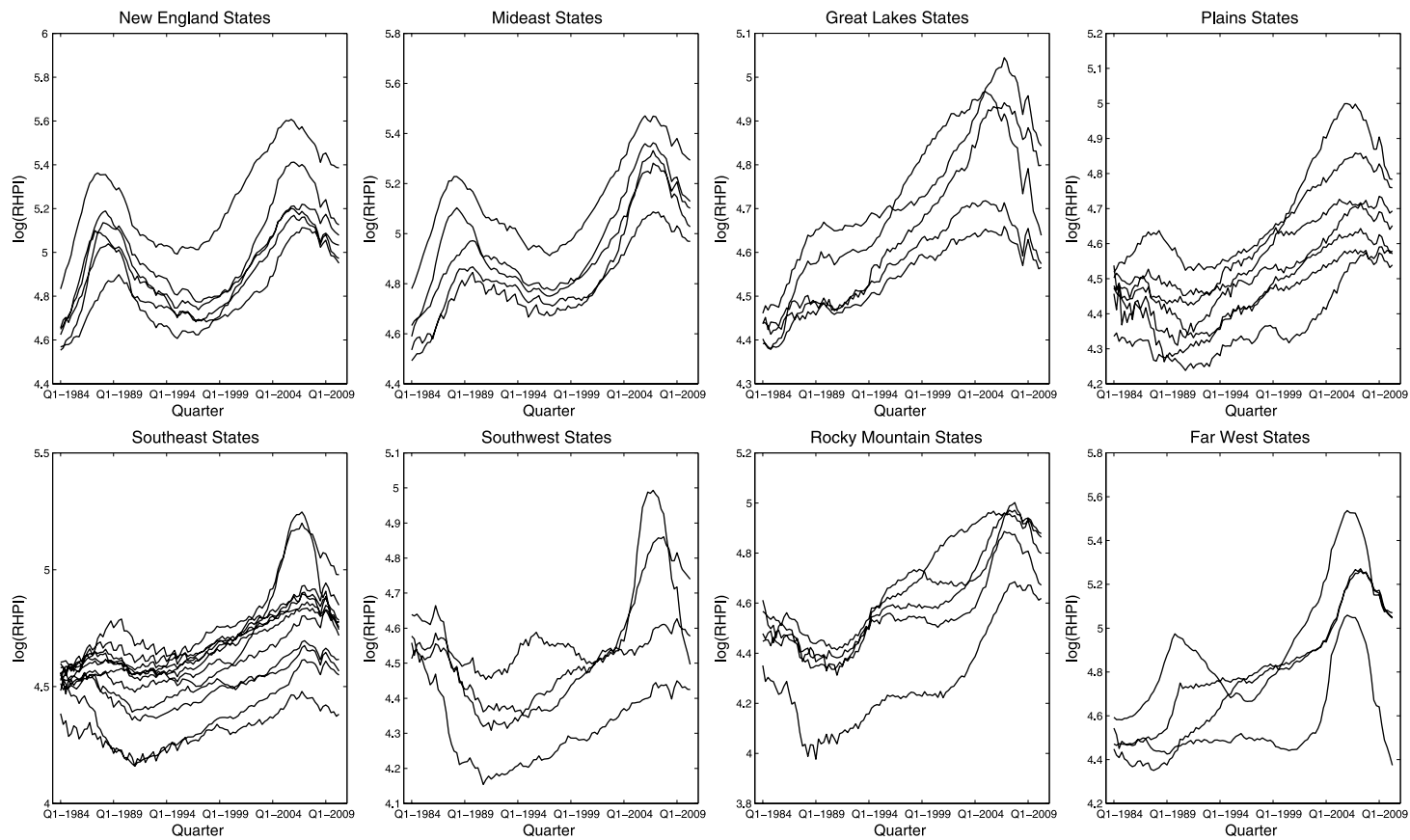}

\caption{Time series of the log-transformed real housing price
index. The 48 United States are grouped in the eight Bureau of Economic
Analysis (BEA) regions.}
\label{figbea}
\end{sidewaysfigure}

The data analyzed in this paper are from the St. Louis Federal Reserve
Bank database\footnote{\url{http://research.stlouisfed.org/fred2/}.} and the
Bureau of Labor Statistics\footnote{\url{http://stats.bls.gov/cpi/home.htm\#data}.} and consist of quarterly time series on $48$ States (excluding
Alaska and Hawaii) from 1984 (first quarter) to 2011 (fourth quarter).
Figure \ref{figbea} shows the time series of the real housing price
index for the 48 United States grouped in the eight Bureau of Economic
Analysis (BEA) regions. The time series are expressed in a logarithmic
scale---see Section \ref{secapplication} for a complete description of
the data set.

Figure \ref{figbea} shows that there are interesting dynamic
structures in the time series and that periodic patterns and common
trend components are consistent features of the housing market.
Specifically, it appears that housing prices have been rising rapidly.
Since 1995 we have estimated that, on average, real housing prices have
increased about 36 percent, roughly double the increase of previous
housing price booms observed in the late 1980s. Moreover, we notice
that housing prices continued to rise strongly during the 2001
recession and that the process of the housing price boom, which some
have interpreted as a bubble, started in 1998, accelerated during the
period 2003--2006 and burst in 2007. The prices have then been falling
sharply over all the country.

The possibility of modeling all these dynamic features, as well as to
obtain accurate housing price forecasts, is important for prospective
homeowners, investors, appraisers and other real estate market
participants, such as mortgage lenders and insurers.

\begin{table}
\caption{Average of correlation coefficients within and between
regions first difference log of real housing prices. BEA regions:
New England (NE), Mideast (ME), Great Lakes (GL), Plains (PL),
Southeast (SE), Southwest (SW), Rocky Mountain (RM), Far West (FW)}
\label{tabcorr}
\begin{tabular*}{\textwidth}{@{\extracolsep{\fill}}lcccccccc@{}}
\hline
& \textbf{NE} & \textbf{ME} & \textbf{GL} & \textbf{PL} & \textbf{SE} & \textbf{SW} & \textbf{RM} & \textbf{FW} \\
\hline
NE & 0.80 & -- & -- & -- & -- & -- & -- & -- \\
ME & 0.72 & 0.74 & -- & -- & -- & -- & -- & -- \\
GL & 0.47 & 0.48 & 0.63 & -- & -- & -- & -- & -- \\
PL & 0.23 & 0.25 & 0.35 & 0.50 & -- & -- & -- & -- \\
SE & 0.35 & 0.40 & 0.48 & 0.36 & 0.45 & -- & -- & -- \\
SW & 0.24 & 0.29 & 0.35 & 0.42 & 0.42 & 0.47 & -- & -- \\
RM & 0.10 & 0.17 & 0.30 & 0.46 & 0.37 & 0.48 & 0.50 & -- \\
FW & 0.33 & 0.46 & 0.42 & 0.34 & 0.37 & 0.40 & 0.41 & 0.50 \\
\hline
\end{tabular*}
\end{table}

The way in which housing prices spread out to surrounding locations
over time are also of interest in the real estate literature. The
co-movements shown by the time series within BEA regions suggest the
presence of spatial correlation. As stated in \citet{HolPesYam10}, it is possible that States that are contiguous may
influence each other's housing prices. In fact, high prices in
metropolitan areas may persuade people to commute from neighboring
States. Labour mobility is quite high in the USA and lower housing
prices may provide an incentive to migrate. Another possible source of
cross-sectional dependence would be due to economy-wide common shocks
that affect all cross section units. Changes in interest rates, oil
prices and technology are examples of such common shocks that may
affect housing prices, although with different degrees across States.

To explore the existence of spatial interactions, using data on the
growth of real housing prices, Table \ref{tabcorr} shows the simple
correlation coefficients between each State, within and between
correlations for the $8$ BEA regions. The diagonal elements show the
within region average correlation coefficients, while the off-diagonal
elements give the between region correlation coefficients.
Apart from the States of the Southeast, which are more correlated on
average with the States of the Great Lakes than among themselves,
the within region correlation is larger than the between region
correlation. In general, on average, the correlations decline with
distance, but it is interesting to note the quite high correlations
between the East and West regions, that is, for States belonging to the
Mideast and Far West regions. In general, there is more evidence in the
raw data of a possible spatial pattern in real housing prices than in
real incomes and unemployment rate.

\subsection{Related literature and the proposed model}

Modeling the spatio-\break temporal variability of housing prices has enjoyed
widespread popularity in the last years. In order to obtain a high
degree of accuracy in the results, the analysis of housing prices
across US States requires the definition of a general and flexible
econometric model where the temporal and cross-sectional dependencies
must be accommodated. Several efforts have been made to develop
spatio-temporal models but there is no single approach which can be
considered uniformly as being the most appropriate. For example, time
series models have become increasingly sophisticated in their treatment
of dynamics and trends over time, including the application of unit
roots and cointegration techniques [\citet{GiuHad91},
\citet{Mee01}, \citet{MueMur97}]. However, traditional
approaches, such as those based on standard vector autoregression
analysis (VAR), do not allow for a direct modeling of locational
spillovers and are thus not consistent with the ``ripple effect'' theory
[\citet{Mee99}]. A spatial adaptation of VARs, denoted as SpVAR models,
explicitly considers the potential impacts of economic events in
neighboring States and has been discussed in \citet{KuePed11}.
The SpVAR is a specific version of the Spatio-Temporal Auto-Regressive
Moving Average---(STARMA)---model introduced by \citet{PfeDeu80} where the linear dependencies are lagged in both space and time.
Since STARMAs are an extension of the ARMA class of models [\citet{BoxJenRei94}], they are particularly useful to produce
temporal forecasts of the variable of interest. However, the STARMA
specification also suffers from some disadvantages. First, because of
the amount of computational effort required, STARMAs are in general
only suitable for modeling data which are dense in time and sparse in
space. For example, in \citet{KuePed11} the analysis is only
limited to $11$ States (i.e., West Region). Secondly, the understanding
of co-movements among US State housing prices (and other involved
variables) is difficult when the number of the States is large.
Knowledge of this covariation is required both to academics seeking to
explain the economic nature and sources of variation and to
practitioners involved in the development of trading strategies.
Thirdly, as argued by Anselin [(\citeyear{Ans88}), pages~11--14], the STARMA class
does not offer a fully adequate modeling of the spatial dependence and
heterogeneity of observations. The lack of an adequate treatment of a
simultaneous (instantaneous) spatial dependence is also the main point
of criticism raised by Cressie [(\citeyear{Cre93}), page 450] to the STARMA
methodology. In fact, in its standard specification, STARMA implicitly
assumes that, conditional on past observations, the process is
uncorrelated across space. This is undoubtedly a major shortcoming,
since many observed series, as noted, for example, by \citet{PfeDeu81}, show considerable contemporaneous correlation even
after conditioning on the past history of the process. When the
contemporaneous correlation is considered by the model, the
observations become a nonlinear transformation of the innovations and,
as a result, maximum likelihood estimation becomes much more difficult
[\citet{Elh01}, \citet{DiGetal05}].

Seemingly Unrelated Regression (SUR) and error correction panel data
models [see, e.g., \citet{Mee01}, Cameron, Muellbauer and Murphy
(\citeyear{CamMueMur06})] have also been largely used with spatial and time effects to
investigate the evolution of housing prices. Apart from their rather
complex structure, as STARMAs, these models are not suitable when the
number of regions is relatively large. In fact, the application of an
unrestricted SURE-GLS approach to large $N$ (cross section dimension)
and~$T$ (time series dimension) panels involves nuisance parameters
that increase at a quadratic rate as the cross section dimension of the
panel is allowed to rise [\citet{Pes06}].

Recent research has found that in a data rich environment, dimension
reduction in the form of factors is useful for exploratory analysis,
prediction and policy analysis. Factor analysis assumes that the cross
dependence can be \mbox{characterized} by a finite number of unobserved common
factors, possibly due to economy-wide shocks that affect all States,
albeit with different intensities. Thus, strong co-movement and high
correlation among the series suggest that both observable and
unobservable factors must be at place. The effects of common shocks on
housing prices have been taken in consideration in \citet{vanetal11} and Holly, Pesaran and
Yamagata (\citeyear{HolPesYam10}) by making use of the common correlated
effects estimator [CCE, \citet{Pes06}] which controls for
heterogeneity and spatial dependence. In these studies, the authors
develop a panel data model where fixed mean effects, cointegration,
cross-equation correlations and latent factors are considered.
Furthermore, they show that by approximating the linear combinations of
the unobserved factors by cross section averages of the dependent and
explanatory variables, and by running standard panel regressions
augmented with these cross section averages, spatial dependency can be
eliminated.

Differently from these authors, we approach the analysis from the
perspective of recent developments of dynamic factor models in the
literature of spatio-temporal processes. We assume that the observed
process can be modeled by a temporally dynamic and spatially
descriptive model, hereafter referred to as the \textit{spatial dynamic
structural equation} model---SD-SEM. There are some important
differences between our approach and the one discussed by Holly, Pesaran and
Yamagata
(\citeyear{HolPesYam10}) and \citet{vanetal11}. Firstly, differently from these
authors, we do not use cross section averages to eliminate
cross-sectional dependencies. Instead, our model formulation exploits
the spatio-temporal nature of the data and explicitly defines a
nonseparable spatio-temporal covariance structure of the multivariate
process. Secondly, because of the high dimensionality of the data,
dimension reduction is important and we suggest modeling the temporal
relationship between dependent and regressor variables in a latent
space. The observed processes are thus described by a potentially small
set of common dynamic latent factors. For all possible model candidates
which may be specified, we use a multivariate autoregressive
distributed-lag specification for these latent processes and, to
account for situations in which two or more latent factors appear to
exhibit a common trend, their cointegrating relationship is considered.
Thirdly, by modeling the spatial variation via spatially structured
factor loadings, we entertain the possibility of identifying clusters
of States that share common time series components. This is one of the
main advantages of our model formulation. Lastly, the model naturally
allows for producing temporal and spatial predictions of the variables
of interest. Note that although spatial interpolation is not a main
task in lattice data applications, it may be an important issue in
terms of missing data reconstruction (i.e., partial or total
reconstruction of the housing price time series). This problem would
not be easily addressed by the other model formulations discussed above.

The SD-SEM represents a multivariate extension of the model recently
proposed by Ippoliti, Gamerman and Valentini (\citeyear{IppValGam12}) for modeling
environmental coupled (correlated) spatio-temporal processes. Our
spatio-temporal data are thus multivariate, in that more than one
variable is typically measured at specific spatial sites (States) and
different temporal instants. Furthermore, as in \citet{LopSalGam08} and Ippoliti, Valentini and
Gamerman (\citeyear{IppValGam12}), we assume that the spatial
dependence can be modeled through the columns of the factor loading
matrices. However, differently from these authors, who refer to
applications with spatially continuous (i.e., geostatistical)
processes, we consider here applications with lattice data such that
the factor loadings can be modeled as conditionally independent
multivariate Gaussian Markov Random Fields---GMRFs. While models for
multivariate geostatistical data have been extensively explored, models
for lattice data have received less attention in the literature. For
recent methodological developments the reader is referred to \citet{SaiCre07}, \citet{SaiFurCre11} and the references therein.

The SD-SEM is developed within a state-space framework and full
probabilistic inference for the parameters is facilitated by Markov
chain Monte Carlo (MCMC).

The remainder of the paper is organized as follows. In Section \ref
{secDLM1} we describe the general dynamic latent model, while in
Section \ref{secpatterns} specific attention is given to models which
incorporate general forms of the spatial correlations and
cross-correlations between variables at different locations. In Section
\ref{secstatespace} we describe the state-space formulation and in
Section \ref{secnonstationary} discuss the nonstationary cases for the
temporal dynamics of the latent factors. In Section \ref{secinference}
we consider Bayesian inferential issues and in Section \ref
{secforecasting} we describe forecasting strategies. In Section \ref
{secapplication} we discuss fits of the model to the data set of US
real housing prices, while Section \ref{secdiscussion} concludes the paper.

\section{The spatial dynamic structural equation model}\label{secDLM1}
Often observations are multivariate in nature, that is, we obtain
vector responses at locations across space. For such data,
we need to model both association between measurements at a location as
well as association between measurements across locations.
With increased collection of such multivariate spatial data, there
arises the need for flexible explanatory stochastic models in order to
improve estimation precision [see, e.g., \citet{KimSunTsu01}] and to provide simple descriptions of the complex relationships
existing among the variables. In the following, a model formulation
which describes the structural relations among the variables
in a lower dimensional space is presented.

Assume that $Y$ and $X$ are two multivariate (multidimensional)
spatio-temporal processes, that is, assume that several variables are
measured at the node or interior (State), $\mathbf{s}$, of a lattice
$\mathcal{L}$ and temporal instant $t\in\{1,2,\ldots,T\}$. Hence, for
$n_y$ variables, we write $\mathbf{Y}(\mathbf{s},t)=[Y_1(\mathbf{s},t),
\ldots, Y_{n_y}(\mathbf{s},t)]'$, and the same holds for $X$, for
$n_x$ variables.
It is explicitly assumed that ${X}$ is a predictor of ${Y}$, which is
the process of interest.

Also, assume that $N$ is the number of locations in $\mathcal{L}$ and
let $\tilde{n}_y =n_y N$ and $\tilde{n}_x =n_x N$. Then, at a specific
time $t$, the $(\tilde{n}_y\times1)$ and $(\tilde{n}_x\times1)$
dimensional spatial processes, $Y$ and $X$, are denoted as $\mathbf
{Y}(t) = [\mathbf{Y}(\mathbf{s}_1,t)', \ldots, \mathbf{Y}(\mathbf
{s}_N,t)' ]'$ and $\mathbf{X}(t) = [\mathbf{X}(\mathbf{s}_1,t)', \ldots, \mathbf{X}(\mathbf{s}_N,t)' ]'$.

Our model assumes that each multivariate spatial process, at a specific
time $t$, has the following linear structure:
%
\begin{eqnarray}
\mathbf{X}(t)& = & \mathbf{m}_x(t)+ \mathbf{H}_x
\mathbf{f}(t) + \mathbf{u}_x(t), \label{eqmeasx}
\\
\mathbf{Y}(t)& = & \mathbf{m}_y(t)+ \mathbf{H}_y
\mathbf{g}(t) + \mathbf{u}_y(t), \label{eqmeasy}
\end{eqnarray}
where $\mathbf{m}_y(t)$ and $\mathbf{m}_x(t)$ are $(\tilde
{n}_y \times1)$ and $(\tilde{n}_x\times1)$ mean components modeling
the smooth large-scale temporal variability, $\mathbf{H}_y$ and $\mathbf
{H}_x$ are measurement (factor loadings) matrices of dimensions $(
\tilde{n}_y \times m)$ and $( \tilde{n}_x \times l)$, respectively, and
$\mathbf{g}(t)$ and $\mathbf{f}(t)$ are $m$- and $l$-dimensional
vectors of temporal common factors.
Also, $\mathbf{u}_y(t)$ and $\mathbf{u}_x(t)$ are Gaussian error terms
for which we assume $\mathbf{u}_y(t) \sim N(\mathbf{0},\bolds{\Sigma
}_{u_y})$ and $\mathbf{u}_x(t) \sim N(\mathbf{0},\bolds{\Sigma
}_{u_x})$. For simplicity, throughout the paper it is assumed that
$\bolds{\Sigma}_{u_y}$ and $\bolds{\Sigma}_{u_x}$ are both diagonal
matrices and that $m\ll\tilde{n}_y$ and $l \ll \tilde{n}_x$.

The temporal dynamic of the common factors is then modeled through the
following state equations:
%
\begin{eqnarray}
\mathbf{g}(t)& = & \sum_{i=1}^{p}
\mathbf{C}_i \mathbf{g}(t-i)+ \sum_{j=1}^{q}
\mathbf{D}_j \mathbf{f}(t-j)+\boldxi(t), \label {eqstateg}
\\
\mathbf{f}(t)& = & \sum_{k=1}^{s}
\mathbf{R}_k \mathbf{f}(t-k)+ \boldeta(t), \label{eqstateff}
\end{eqnarray}
where $\mathbf{C}_i$ $(m \times m)$, $\mathbf{D}_j$ $(m
\times l)$, and $\mathbf{R}_k$ $(l \times l)$ are coefficient matrices
modeling the temporal evolution of the latent vectors $\mathbf
{g}(t)= [g_1(t), \ldots, g_m(t) ]'$ and $\mathbf{f}(t)=
[f_1(t), \ldots,f_l(t) ]'$, respectively. Finally, $\boldxi(t)$
and $\boldeta(t)$ are independent Gaussian error terms for which we
assume $\boldxi(t) \sim N(\mathbf{0},\bolds{\Sigma}_\xi)$ and $\boldeta
(t) \sim N(\mathbf{0},\bolds{\Sigma}_\eta)$.

Equation (\ref{eqstateg}) represents a Vector Autoregressive model
with exogenous variables (VARX) where the variables in $\mathbf{g}(t)$,
considered as endogenous (i.e., determined within the system), are
controlled for the effects of other variables, $\mathbf{f}(t)$,
considered as exogenous (i.e., determined outside the system and
treated independently of the other variables)\footnote{The distinction
between ``exogenous'' and ``endogenous'' variables in a model is subtle
and is a subject of a long debate in the literature. See, for example,
\citet{EngHenRic83}, \citet{OsiSte96}.
Gourieroux and Monfort [(\citeyear{GouMon97}), Chapter 10] also provide a clear
distinction between the different exogeneity concepts.}. Equations
(\ref{eqmeasx})--(\ref{eqstateff}) thus provide the basic formulation of the SD-SEM. One advantage
of this model is that temporal forecasts of the variable of interest,
$Y$, can be obtained by modeling the dynamics of a few common factors.
Also, the model is spatially descriptive in that it can be used to
identify possible clusters of locations whose temporal behavior is
primarily described by a potentially small set of common dynamic
latent factors. As it will be shown in the next section, flexible and
spatially structured prior information regarding such clusters can be
specified through the columns of the factor loading matrix.

\section{Factor loadings and multivariate GMRFs}\label{secpatterns}
A key property of much spatio-temporal data is that observations at
nearby sites and times will tend to be similar to one another. This
underlying smoothness characteristic of a space--time process can be
captured by estimating the state process and filtering out the
measurement noise.
It is customary for dynamic latent models to refer to the unobserved
(state) processes as the common factors and to refer to the
coefficients that link the factors with the observed series as the
factor loadings. It is assumed that these factor loadings have the
nature of spatial processes and, extending results in Ippoliti, Valentini and
Gamerman
(\citeyear{IppValGam12}), here the spatial dependence is modeled through a multivariate
GMRF. Relevant papers useful for our purposes are \citet{Mar88} and
\citet{SaiCre07}, and we refer to them for known results on the
model formulation.

Let $\mathbf{h}_{x_j}=[\mathbf{h}_{x_j}(\mathbf{s}_1)^\prime, \mathbf
{h}_{x_j}(\mathbf{s}_2)^\prime, \ldots, \mathbf{h}_{x_j}(\mathbf
{s}_N)^\prime]'$, that is, the $j$th column of $\mathbf{H}_{x}$, be a
$\tilde{n}_x$-dimensional spatial process observed on $\mathcal{L}$---and similarly for $\mathbf{H}_{y}$. Also, let $[\mathbf{h}_{x_j}(\mathbf
{s}_i)| {R}_{-i}]$ denote the conditional distribution of $\mathbf
{h}_{x_j}(\mathbf{s}_i)$ given the rest (i.e., values at all other
sites). Then, the GMRF is defined by the conditional mean
%
\begin{equation}
\label{mediacondiz} \mathrm{E} \bigl(\mathbf{h}_{x_j}(
\mathbf{s}_i)| {R}_{-i} \bigr) = \boldmu
_{i}^{(h_{x_j})}+ \sum_{u \in\mathcal{S}_i}\mathbf
{F}_{iu}^{(h_{x_j})} \bigl(\mathbf{h}_{{x_j}}(
\mathbf{s}_u) - \boldmu _{u}^{(h_{x_j})} \bigr)
\end{equation}
and the conditional covariance matrix
%
\begin{equation}
\label{varianzacondiz} \operatorname{Var} \bigl(\mathbf{h}_{{x_j}}(
\mathbf{s}_i)| {R}_{-i} \bigr) = \mathbf{T}_i^{(h_{x_j})},
\end{equation}
where $\mathcal{S}_i$\vspace*{-1pt} is a finite subset of $\mathcal{L}$
containing neighbors of site $\mathbf{s}_i$, $\boldmu_{i}^{(h_{x_j})}$
is a $n_x$-dimensional mean vector, and $\mathbf{F}_{iu}^{(h_{x_j})}$
is a $(n_x \times n_x)$ matrix of spatial regression parameters.

To take into account the effect of some explanatory variables, it is
possible to parameterize the mean vector through the definition of a
$(N \times q)$ design matrix, ${\bolds{\mathcal{D}}}^*$, such that $\boldmu
^{(h_{x_j})}={\bolds{\mathcal{D}}}^* \boldbeta^{(h_{x_j})} $, with
$\boldbeta^{(h_{x_j})}$ a $(q \times1)$ vector of parameters. Assuming
$\boldc_i$ is a vector of covariates for the $i$th location, we have
$\boldmu^{(h_{x_j})}=  [ \boldmu_{1}^{(h_{x_j})^\prime}, \ldots,
\boldmu_{N}^{(h_{x_j})^\prime}  ]^\prime$, with $\boldmu
_{i}^{(h_{x_j})}=\bolds{\mathcal{D}}^*_i\boldbeta^{(h_{x_j})}_i$,
$\boldbeta^{(h_{x_j})}= [\boldbeta_1^{(h_{x_j})^\prime},\ldots,
\boldbeta_{n_x}^{(h_{x_j})^\prime} ]^\prime$, $\mathcal{\bolds{\mathcal{D}}}^*_i=(\mathbf{I}_{n_x} \otimes\boldc_i^\prime),   i=1,
\ldots, n$, and $\otimes$ denoting the Kronecker product. For a
discussion of different specifications of the matrix $\bolds{\mathcal
{D}}^*$, see, for example, Ippoliti, Valentini and
Gamerman (\citeyear{IppValGam12}) and Lopes, Salazar and
Gamerman
(\citeyear{LopSalGam08}). However, due to the static behavior of $\mathbf{h}_{x_j}$, only
spatially-varying covariates will be
considered in explaining the mean level of the GMRF.

With the definition of the conditional distributions, it follows [see
\citet{Mar88}] that the joint distribution of $\mathbf{h}_{x_j}$ is
$\operatorname{MVN}  (\boldmu^{(h_{x_j})}, \bolds{\Sigma}^{(h_{x_j})} )$ with
the covariance matrix specified as $\bolds{\Sigma}^{(h_{x_j})}=
\{\operatorname{block}  [-\mathbf{T}_{i}^{(h_{x_j})^{-1}}\mathbf
{F}_{iu}^{(h_{x_j})}  ]  \}^{-1}$, where $\mathbf{F}_{ii}= -
\mathbf{I}$ and for a generic matrix $\mathbf{G}$, $\operatorname{block} [\mathbf
{G}_{iu}]$ denotes a block matrix with the $(i,u)$th block given by
$\mathbf{G}_{iu}$ [see \citet{SaiCre07}]. To guarantee that a
proper probability density function is defined, the parametrization
must ensure that $\bolds{\Sigma}^{(h_{x_j})}$ is positive-definite and
symmetric; hence, we require both $\mathbf{F}_{iu}^{(h_{x_j})}\mathbf
{T}_{u}^{(h_{x_j})}=\mathbf{T}_{i}^{(h_{x_j})}\mathbf
{F}_{ui}^{(h_{x_j})^\prime}$ and $\operatorname{block} [-\mathbf
{T}_{i}^{(h_{x_j})^{-1}}\mathbf{F}_{iu}^{(h_{x_j})}  ]$ positive definite.

\section{The state space formulation}\label{secstatespace}
As shown in Section \ref{secDLM1}, the temporal dynamic is
modeled through the state equations (\ref{eqstateg}) and (\ref{eqstateff}).
The specification of equation (\ref{eqstateff}) is necessary to
predict in time the latent process $\mathbf{f}(t)$ and thus to obtain
$k$-step ahead forecasts of $\mathbf{g}(t)$ through equation (\ref
{eqstateg}). It is thus useful to specify the joint generation
process for $\mathbf{g}(t)$ and $\mathbf{f}(t)$ as
%
\begin{eqnarray}
\label{uncpred} \left[\matrix{ \mathbf{g}(t)
\vspace*{2pt}\cr
\mathbf{f}(t) }
\right] &=&\left[ %
\matrix{\mathbf{C}_1 & \mathbf{D}_{1}
\vspace*{2pt}\cr
\mathbf{0} & \mathbf{R}_1 }
\right] \left[\matrix{ %
 \mathbf{g}(t-1)
\vspace*{2pt}\cr
\mathbf{f}(t-1) }
\right] + \cdots
\nonumber
\\[-8pt]
\\[-8pt]
\nonumber
&&{}+ \left[
\matrix{\mathbf{C}_p & \mathbf{D}_{p}
\vspace*{2pt}\cr
\mathbf{0} & \mathbf{R}_p }
\right] \left[ \matrix{ \mathbf{g}(t-p)
\vspace*{2pt}\cr
\mathbf{f}(t-p)}
\right]+ \left[ \matrix{ \boldxi(t)
\vspace*{2pt}\cr
\boldeta(t) }
\right],
\end{eqnarray}
where it is assumed without loss of generality that $p
\geq \operatorname{max}(s,q)$, $\mathbf{D}_{i}=\mathbf{0}$ for $i>q$ and
$\mathbf{R}_{j}=\mathbf{0}$ for $j>s$. It follows that the joint
generation process of
$\mathbf{g}(t)$ and $\mathbf{f}(t)$ is a VAR($p$) process of the type
%
\begin{equation}
\label{varfin} \mathbf{d}(t) = \boldPhi_1 \mathbf{d}(t-1)+ \cdots+
\boldPhi_p \mathbf{d}(t-p) + \boldeps(t),
\end{equation}
where
\[
\mathbf{d}(t)=\left[ %
\matrix{ \mathbf{g}(t)
\vspace*{2pt}\cr
\mathbf{f}(t) }
\right],\qquad \boldPhi_i=
\left[ \matrix{ \mathbf{C}_i &
\mathbf{D}_{i}
\vspace*{2pt}\cr
\mathbf{0} & \mathbf{R}_i}
\right],\qquad
\boldeps(t)=\left[ \matrix{ \boldxi(t)
\vspace*{2pt}\cr
\boldeta(t) }
\right].
\]

The presence of the measurement and the state variables naturally leads
to the state-space representation [Lutkepohl (\citeyear{Lut05})] of the SD-SEM
model; given the data, this representation allows for a recursive
estimate of the latent variables through the Kalman filter algorithm.
The linear Gaussian state-space model is thus described
by the following \textit{state} and \textit{measurement} equations:
\begin{eqnarray}
\boldalpha(t) &= & \bolds{\Phi} \boldalpha(t-1)+ \boldXi \boldzeta(t),
\label{Seq}
\\
\mathbf{z}(t) & = & \mathbf{H} \boldalpha(t) + \mathbf{u}(t), \label{Meq}
\end{eqnarray}
where $\boldalpha(t)$ is the state vector,
$\bolds{\Phi}$ is the nonsingular transition matrix,
$\boldXi$ is a constant input matrix, $\mathbf{z}(t)$ is the
measurement vector and \textbf{H} is the measurement matrix. The
sequences $\boldzeta(t)$ and $\mathbf{u}(t)$ are assumed to be
mutually independent zero mean Gaussian random variables with
covariances $ E\{\boldzeta(t_i)\boldzeta(t_j)'\}=\bolds{\Psi}\delta
_{ij}$ and $ E\{\mathbf{u}(t_i)\mathbf{u}(t_j)'\}=\bolds{\Sigma
}_u\delta_{ij}$, where $E\{\cdot\}$ denotes the expectation and
$\delta_{ij}$ the Kronecker delta function. In (\ref{Seq}) and (\ref
{Meq}) we have the following specification:
\begin{eqnarray*}
\boldalpha(t)&=& \left[ %
\matrix{ \mathbf{d}(t)
\vspace*{2pt}\cr
\mathbf{d}(t-1)
\vspace*{2pt}\cr
\vdots
\vspace*{2pt}\cr
\mathbf{d}(t-p+1)}
\right],\qquad \bolds{\Phi}=\left[
\matrix{\boldPhi_1 & \boldPhi_2
& \cdots& \boldPhi_p
\vspace*{2pt}\cr
\mathbf{I} & \mathbf{0} & \cdots& \mathbf{0}
\vspace*{2pt}\cr
\vdots& \vdots & \vdots& \vdots
\vspace*{2pt}\cr
\mathbf{0} & \cdots & \mathbf{I} & \mathbf{0}}
\right], \\
\boldzeta(t)&=& \left[ %
\matrix{ \boldeps(t)
\vspace*{2pt}\cr
\mathbf{0}
\vspace*{2pt}\cr
\vdots
\vspace*{2pt}\cr
\mathbf{0} }
\right],\qquad
\mathbf{z}(t)=\left[ %
\matrix{ \mathbf{y}(t)
\vspace*{2pt}\cr
\mathbf{x}(t)}
\right],\qquad \mathbf{H}=\left[\matrix{\mathbf{H}_y & \mathbf{0} & \cdots& \mathbf{0}
\vspace*{2pt}\cr
\mathbf{0} & \mathbf{H}_x & \cdots& \mathbf{0}}
\right], \\
\boldXi&=&\left[\matrix{
\mathbf{I}
\vspace*{2pt}\cr
\mathbf{0}
\vspace*{2pt}\cr
\vdots
\vspace*{2pt}\cr
\mathbf{0} }
\right], \qquad
\mathbf{u}(t)= \left[\matrix{
\mathbf{u}_y(t)
\vspace*{2pt}\cr
\mathbf{u}_x(t)}
\right].
\end{eqnarray*}

\section{Nonstationary latent factors}\label{secnonstationary}
The dynamic specification for the state vector $\boldalpha
(t)$ is quite general. In fact, the family of time series processes
that can be formulated as in equations (\ref{Seq}) and (\ref{Meq}) is
wide and includes a broad range of nonstationary time series processes.
Sometimes it may be advantageous to have a specification that
decomposes the latent factors into stationary and nonstationary
components, such as trend, periodic or cyclical components.The large
scale dynamic components can in fact be directly specified through the
common dynamic factors. In this case, for example, common seasonal
factors can receive different weights for different columns of the
factor loading matrix, so allowing different seasonal patterns for the
spatial locations. For some specific examples, and for a wider
discussion on this point, see Lopes, Salazar and
Gamerman (\citeyear{LopSalGam08}) and Ippoliti, Valentini and
Gamerman (\citeyear{IppValGam12}).

\subsection{Cointegrated latent factors}\label{secsubnonstationary}
Nonstationarity can also occur when two or more latent factors appear
to exhibit a common trend, and hence are cointegrated [\citet{Joh88}]. In this case we have that one or more linear combinations of these
factors are stationary even though individually they are not. If the
factors are cointegrated, they cannot move too far away from each other
and we should observe a stable long-run
relationship among their levels. In contrast, a lack of cointegration
suggests that such factors have no long-run link and, in principle,
they can wander arbitrarily far away from each other.

In our model formulation we consider the case in which the exogeneous
factors are cointegrated among themselves as well as with the
endogenous latent variables. In this case the vector autoregressive
process of equation (\ref{varfin}) can be written in the error
correction model (ECM) form as
%
\begin{equation}
\label{eqmodelnostationary} \Delta\mathbf{d}(t) =  \tilde{ \mathbf{A}}
\mathbf{d}(t-1) + \sum_{i=1}^{p-1} \tilde{
\boldPhi}_i \Delta\mathbf{d}(t-i)+\boldeps(t),
\end{equation}
where $\tilde{ \mathbf{A}}=-\mathbf{I} + \sum_{i=1}^{p}
\boldPhi_i $, $\tilde{\boldPhi}_i = - \sum_{j=i+1}^{p} \boldPhi_j $
and $\Delta$\vspace*{1pt} is the difference operator, that is, $\Delta\mathbf{d}(t)
= \mathbf{d}(t) - \mathbf{d}(t-1)$. Full details of the vector error
correction specification of equation (\ref{eqmodelnostationary}) are
provided in Appendix \ref{appcoint} where we also show that the matrix
of long-run multipliers, $\tilde{\mathbf{A}}$, is an upper block
triangular matrix. These single blocks, expressed as a product of
parameter matrices, provide information about: (i) the cointegration
structure \textit{within} the exogenous and endogenous processes
$\mathbf{f}(t)$ and $\mathbf{g}(t)$, and (ii) the cointegration \textit
{between} the two processes.

\section{Inference and computations}\label{secinference}

\subsection{Prior information}
Full probabilistic inference for the model parameters is carried out
based on the following independent prior distributions. Throughout we
shall use $\operatorname{vec}(\cdot)$ to denote the vec operator and $G(a,b)$ to
denote the Gamma distribution with mean $a/b$ and variance $a/b^2$.
Unless explicitly needed, full specifications of the priors are only
given for $X$ so that definitions for $Y$ follow accordingly.

\textsc{Measurement equation}.
The precision matrix $\bolds{\Sigma}_{u_x}^{-1}$ is assumed
to be diagonal where each element has a Gamma prior distribution,
$G(0.01,0.01)$.

The prior distribution for $\boldbeta^{(h_{x_i})}$ ($i=1,\ldots,l$) is
$N(\mathbf{0}, \sigma_\beta^2 \mathbf{I})$. Then, assuming a constant
conditional covariance matrix, the prior on the inverse covariance
matrix $\mathbf{T}^{(h_{x_i})^{-1}}$ is given by the Wishart
distribution [\citet{MarKenBib79}], that is, $\mathbf
{T}^{(h_{x_i})^{-1}} \sim W  (\varrho_x, (\varrho_x \mathbf
{S}_x)^{-1}  )$, where $\varrho_x > l$ and $\mathbf{S}_x$ is a
pre-specified symmetric positive definite matrix. To provide the prior
specification for the joint distribution of the spatial regression
parameters, we set ${\mathbf{F}}_{iu}^{(h_{x_i})}={\mathbf
{F}}^{(h_{x_i})}$ and, following \citet{SaiCre07}, we use the
reparametrization $\tilde{\mathbf{F}}^{(h_{x_i})}=\mathbf
{T}^{(h_{x_i})^{-1/2}}\mathbf{F}^{(h_{x_i})}\mathbf
{T}^{(h_{x_i})^{1/2}}$ and specify its prior to be proportional to
$\exp \{ -\boldupsilon' \boldupsilon/\varsigma^2  \}$, where
$\boldupsilon= \operatorname{vec} (\tilde{\mathbf{F}}^{(h_{x_i})^\prime}  )$.
The prior parameter $\varsigma$ is specified by choosing small values,
since the prior for $\tilde{\mathbf{F}}^{(h_{x_i})}$ is concentrated
around zero. Then, in both mean and variance of the GMRF processes we
adopt priors centered around prefixed values, as defined in Section~\ref{secpatterns}.

\textsc{State equation}.
When stationarity conditions are met for the latent processes
the prior distributions for the state equation coefficients can be
specified as proposed in Lopes, Salazar and
Gamerman (\citeyear{LopSalGam08}).
For the cointegration case, since the formulation given in equation
(\ref{eqmodelnostationary}) is quite general, and many plausible
restricted models can be envisaged, Stochastic Search Variable
Selection (SSVS) priors [see \citet{Jocetal11}] are used for the
parameters of the state equations. Note that these plausible models may
differ in the choice of the restrictions on the cointegration space,
the number of exogenous and endogenous latent variables, and the lag
length allowed for the autoregression.

The error covariance matrices are assumed to be decomposed as
$\boldSigma_\xi^{-1}=\mathbf{V}_\xi\mathbf{V}_\xi'$ and $\boldSigma_\eta
^{-1}=\mathbf{V}_\eta\mathbf{V}_\eta'$, where $\mathbf{V}_\xi$ and
$\mathbf{V}_\eta$ are upper-triangular matrices. Then, the SSVS priors
involve using a standard Gamma prior for the square of each of the
diagonal elements of $\mathbf{V}_{(\cdot)}$ and the SSVS mixture of
normals prior for each element above the diagonal [\citet{GeoSunNi08}].
Note that if the error covariance matrices are chosen to be
diagonal, then the computation of the posterior simplifies considerably.

Since $\tilde{\mathbf{A}}$ is potentially of reduced rank and crucial
issues of identification may arise in the ECM form, linear identifying
restrictions are usually imposed. However, because of local
identifiability problems and the restriction on the estimable region of
the cointegrating space [\citet{Kooetal06}], the so-called \emph
{linear normalization} approach also suffers from several drawbacks. To
overcome these problems, we thus adopt the SSVS approach proposed by
\citet{Jocetal11} which, defining priors on the cointegration
space, is facilitated by the computation of Gaussian posterior\vadjust{\goodbreak}
conditional distributions [Koop, Leon-Gonzalez and Strachan (\citeyear{KooLeoStr10})].
A~brief summary of the SSVS priors used in this paper is provided in
Appendix \ref{appSSVSprior}. For a more complete description, the
reader is referred to \citet{Jocetal11} and Koop, Le{\'o}n-Gonz{\'a}lez and
Strachan (\citeyear{KooLeoStr10}).

Finally, the prior for the latent process $\boldalpha(t)$ is provided
by the transition equation and is completed by $\boldalpha(0) \sim
N(\mathbf{a}_0,\boldSigma_{\alpha0})$, for known hyperparameters
$\mathbf{a}_0$ and $\boldSigma_{\alpha0}$ [\citet{DurKoo01},
\citet{Ros73}].

\subsection{The likelihood function}\label{subseclikelihhod}
To specify the likelihood function, without loss of generality, it will
be assumed that $\mathbf{m}_y(t)=\mathbf{0}$ and $\mathbf
{m}_x(t)=\mathbf{0}$. Conditional on $\boldalpha(t)$, for $t=1, \ldots,
T$, the SD-SEM model can be rewritten as $\mathbf{Z} = \boldalpha
\mathbf{H}'+ \mathbf{U}$, where $\mathbf{Z}= [\mathbf{z}(1), \ldots, \mathbf{z}(T)  ]'$ and $\boldalpha= [\boldalpha(1), \ldots,
\boldalpha(T)  ]'$.
The error matrix, $\mathbf{U}$, is of dimension $(T \times n)$, where
$n=\tilde{n}_x+\tilde{n}_y$, and follows a matrix-variate normal
distribution, that is, $\mathbf{U} \sim N(\mathbf{0},\mathbf
{I}_T,\bolds{\Sigma}_u)$---see \citet{Daw81} and \citet{BroVanFea98}.
Then the deviance, minus twice the log-likelihood is
\begin{eqnarray*}
&&\mathfrak{D}(\mathbf{z}|\bolds{\Theta}, \bolds{\Sigma}_u,
\mathbf{H}, \boldalpha, m, l) \\
&&\qquad= Tn \log(2 \pi) + T \log|\bolds{
\Sigma}_u| + \operatorname{trace} \bigl\{\bolds{\Sigma}_u^{-1}
\bigl(\mathbf{Z} - \boldalpha\mathbf {H}'\bigr)' \bigl(
\mathbf{Z} - \boldalpha\mathbf{H}'\bigr) \bigr\},
\end{eqnarray*}
where $\bolds{\Theta}$ is the full set of model parameters.

\subsection{Posterior inference}
Posterior inference for the proposed class of spatial dynamic factor
models is facilitated
by MCMC algorithms. Standard MCMC for dynamic linear models are adapted
to our model specification such that,
conditional on $l$ and $m$, posterior and predictive analysis are
readily available. In the following, we provide some information on
the relevant conditional distributions. By denoting with ``$u$'' the
suffix for the unobserved data, posterior inference is based on
summarizing the joint posterior distribution $p(\mathbf{Z}^u, \bolds
{\Theta},\boldalpha(0), \boldalpha|\mathbf{Z}).$

The common factors are jointly sampled by means of the well-known
forward filtering backward sampling (FFBS) algorithm [\citet{CarKoh94}, Fr\"{u}hwirth-Schnatter (\citeyear{Fru94})]. All other full conditional
distributions are ``standard'' multivariate Gaussian or Gamma
distributions. An exception is for the spatial parameter matrices,
$\tilde{\mathbf{F}}^{(h_{y_i})}$ and $\tilde{\mathbf{F}}^{(h_{x_i})}$,
and the covariance matrices, $\mathbf{T}^{(h_{y_i})^{-1}}$ and $\mathbf
{T}^{(h_{x_i})^{-1}}$, which are sampled using a Metropolis--Hastings
step. Specific details for the implementation of the full conditional
distributions can be found in Lopes, Salazar and
Gamerman (\citeyear{LopSalGam08}), \citet{SaiCre07} and \citet{Jocetal11}.

\subsection{Model identification}
Some restrictions on $\mathbf{H}_y$ and $\mathbf{H}_x$ are
needed to define a unique model free from identification problems.
Several possibilities can be considered and the solution adopted here
is to constrain the measurement matrices so that they are lower
triangular, assumed to be of full rank. We note here that we have
proper but quantitatively vague priors which can lead to posteriors
that are computationally indistinguishable from improper ones with the
consequence of an MCMC convergence failure. Hence, to avoid relying so
strongly on the prior specification, we prefer to focus on models which
are identified in a frequentist sense. The approach is fully discussed
in Ippoliti, Valentini and
Gamerman (\citeyear{IppValGam12}) and \citet{Stretal11}.

A critical comment to be borne in mind is that the chosen order of the
univariate time series in
the measurement vector influences interpretation of the factors and may
impact on model fitting and assessment, the interpretation
of factors if such is desired, and the choice of the number of factors.
In such cases, the ordering becomes a modeling decision to
be made on substantive grounds, rather than an empirical matter to be
addressed on the basis of model fit. However, from the
viewpoint of forecasting the ordering is irrelevant. For a detailed
discussion on these points see, for example, \citet{LopWes04}.

\subsection{Model selection} \label{secmodelselection}
With this class of model, an important issue is the selection of $m$
and $l$. Several Bayesian selection methods have been developed and for
a discussion, see, for example, Section 4.1 in Lopes, Salazar and
Gamerman (\citeyear{LopSalGam08}).
Here, we consider a simple approach which only considers the variable
of interest, $Y$, and that consists in the minimization of the
following predictive model choice statistic [PMCC, \citet{GelGho98}]:
\[
\operatorname{PMCC}=\frac{\zeta}{\zeta+1} G + P,
\]
where, for our proposed model, $G=\sum_{i,t} (\mathbf
{Y}(\mathbf{s}_i,t)- E[\mathbf{Y}(\mathbf{s}_i,t)_{\mathrm{rep}}])^2$ and $P=\sum_{i,t} \operatorname{Var}[\mathbf{Y}(\mathbf{s}_i,t)_{\mathrm{rep}}].$

This statistic is based on replicates, $\mathbf{Y}(\mathbf
{s},t)_{\mathrm{rep}}$, of the observed data and the summation is taken over
$i=1,\ldots, N$, and $t=1,\ldots, T$. Essentially, the PMCC quantifies
the fit of the model by comparing features of the posterior predictive
distribution, $p(\mathbf{Y}(\mathbf{s},t)_{\mathrm{rep}}|\mathbf{Y}(\mathbf
{s},t))$, to equivalent features of the observed data. The quantity~$G$
is a measure of goodness of fit while $P$ is a penalty term. As the
models become increasingly complex the goodness-of-fit term will
decrease but the
penalty term will begin to increase. Overfitting of model results in
large predictive variances and large values of the penalty function.
The choice of $\zeta$ determines how much weight is placed on the
goodness-of-fit term relative to the penalty term. As $\zeta$ goes to infinity,
equal weight is placed on these two terms. Banerjee, Carlin and Gelfand
(\citeyear{BanCarGel04}) mention that ordering of models is typically insensitive to the
choice of $\zeta$, therefore, we fix $\zeta=\infty$. Notice that at
each iteration of the MCMC we can obtain replicates of the observations
given the sampled values of the parameters.

\section{Uses of the model}\label{secforecasting}
In this section we provide specific details on how to obtain temporal
forecasts of the variable of interest $Y$.

\subsection{Unconditional forecasting}
Temporal forecasts of the variable $Y$ are directly obtained through
the state space formulation of the model. In fact, it is easy to show
that since $\boldalpha(t)|\boldalpha(t-1) \sim N(\boldPhi\boldalpha
(t-1), \boldSigma_\alpha),$ the $k$-step ahead forecast for the dynamic
factors is given by $p(\boldalpha(t+k)|\bolds{\Theta}) \sim N(\boldPhi
^{(k)} \boldalpha(t), \bolds{\Omega}^{(k)}),$ where $\bolds{\Omega
}^{(k)}=\sum_{j=1}^k \boldPhi^{(k-j)} \boldSigma_\alpha\boldPhi
^{(k-j)'}.$ Therefore, the $k$-step ahead predictive density, $p
(\mathbf{z}(t+k)|\mathbf{Z} )$, of the joint process $\mathbf
{Z}=[{Y}   {X}]$ is given by
\begin{eqnarray*}
p \bigl(\mathbf{z}(t+k)|\mathbf{Z} \bigr)&=&\int p \bigl(\mathbf {z}(t+k)|
\boldalpha(t+k), \mathbf{H},\bolds{\Theta} \bigr) p \bigl(\boldalpha(t+k)|
\boldalpha(t), \mathbf{H}, \bolds{\Theta} \bigr)
\\
& &\hspace*{10pt} {}\times p \bigl(\boldalpha(t),\mathbf{H}, \bolds{\Theta}|\mathbf {Z} \bigr) \,d
\boldalpha(t+k) \,d \boldalpha(T) \,d \mathbf{H} \,d\bolds {\Theta}.
\end{eqnarray*}
Draws from $p (\mathbf{z}(t+k)|\mathbf{Z} )$ can be obtained
in three steps. Firstly, $\bolds{\Theta}$ is sampled from its joint
posterior distribution via MCMC. Secondly, conditionally on $\bolds
{\Theta}$, the common factors $\boldalpha(t+k)$ are independent of
$\mathbf{Z}$ and can be sampled from $p(\boldalpha(t+k)|\bolds{\Theta
})$. Thirdly, $\mathbf{z}(t+k)$ is sampled from $p(\mathbf
{z}(t+k)|\boldalpha(t+k), \mathbf{H}, \bolds{\Theta}).$

\subsection{Conditional forecasting}
The forecasting procedure described above is obtained under the
hypothesis that the predictor $X$ is unknown for the period of
interest. However,
quite flexible forecasts can also be obtained conditional on the
potential future paths of specified variables in the model. In fact, it
may happen that some of the future values of certain variables are
known, because data on these variables
are released earlier than data on the other variables. By incorporating
the knowledge of the future path of the $X$ variable, in principle, it
should be possible to obtain more reliable forecasts of $Y$.

Another use of conditional forecasting is the generation of forecasts
conditional on different ``policy/exploratory'' scenarios. These
scenario-based conditional forecasts allow
one to answer the question: \textit{if something happens to $X$ in the
future, how will it affect forecasts of $Y$ in the future?} Hence, a
plurality of plausible alternative futures for $X$ can be considered
and temporal forecasts of $\mathbf{g}(t)$ can be produced conditional
on a specific path of $\mathbf{f}(t)$. Under these assumptions, in the
following, we propose a simple procedure to obtain $\mathbf{g}(T+k)$
given $\mathbf{f}(T+1),\ldots,\mathbf{f}(T+k)$, and all present and
past information, thus avoiding the use of equation (\ref{eqstateff})
to obtain $k$-step ahead forecasts of $\mathbf{f}(t)$.

Suppose that for the period $T+1, T+2, \ldots, T+k$, $X$ is known (or
fixed {a priori}) and that $\mathbf{X}_k= [\mathbf{x}(T+1),
\mathbf{x}(T+2), \mathbf{x}(T+k) ]$. Then, $k$-step ahead
forecasts of $\mathbf{g}(t)$ may be obtained conditional on $\mathbf
{f}_k=[\mathbf{f}(T+1), \mathbf{f}(T+2), \ldots, \mathbf{f}(T+k)]$,
where $\mathbf{f}_k=\mathbf{H}_x^\dag\mathbf{X}_k$ and $\mathbf
{H}_x^\dag$ is the Moore--Penrose pseudo-inverse of $\mathbf{H}_x$.

Finally, note that although spatial interpolation is not a main task in
lattice data applications, the reconstruction of missing data (i.e.,
partial or total reconstruction of the multivariate time series of one---or more---State)
is an important issue in general. This can be simply
done by exploiting the conditional expectation of the GMRF and
following Section 6.2 in Ippoliti, Valentini and
Gamerman (\citeyear{IppValGam12}).

\section{Spatio-temporal analysis of US housing prices}\label{secapplication}
Public policy interventions in housing markets are widespread and a key
question is the extent to which these policies achieve their desired
objectives and whether there are any unintended consequences.
Especially for its relationship with mortgage behavior, in recent
years, real housing prices have been of great concern for many
financial institutions. Understanding the impact of specific factors on
real housing prices is thus of great interest for governments, real
estate developers and investors. In this paper, we examine if the total
personal income (TPI) and the unemployment rate (UR) have some impact
on the housing price index (HPI). The data, introduced in Section \ref
{subsecdata}, consist of quarterly time series on $48$ States
(excluding Alaska and Hawaii) from 1984 (first quarter) to 2011 (fourth
quarter). However, in this study, the last 10 quarters have been
excluded from the estimation procedure and used only for forecast purposes.

In order to consider per capita personal income (PCI), the annual
population series (U.S. Census Bureau) is converted into a quarterly
series through geometric interpolation. Moreover, we consider real per
capita personal income (RPCI) and housing price index (RHPI) dividing
PCI and HPI by a State level general price index. However, since there
is no US State level consumer price index (CPI), following Holly, Pesaran and
Yamagata
(\citeyear{HolPesYam10}), we have constructed a State level general price index based on
the CPIs of the cities or areas. All the variables are analyzed on a
logarithmic scale. Henceforth, the variables are denoted as $Y=\log
(\mathrm{RHPI})$, $X_1=\log(\mathrm{RPCI})$ and $X_2=\log(\mathrm{UR}).$

\textsc{Model specification}: \textit{Measurement equations}.
To provide a full specification of the inverse covariance
matrix of each factor loading, we make use of a \textit{contiguity} or
\textit{adjacency} matrix $\mathbf{W}$. We assume here that $\mathbf
{W}$ has zero diagonal elements and nonnegative off-diagonal elements
which reflect the dependency between States $\mathbf{s}_i$ and $\mathbf
{s}_j$---that is, the neighborhood set $\mathcal{S}_i$. Hence, to
postulate plausible relationships between two States, as in Holly, Pesaran and
Yamagata (\citeyear{HolPesYam10}), we assume that $\mathbf{W}$ is a binary proximity matrix
which assigns uniform weights to all neighbors of State $\mathbf{s}_i$,
that is,
\[
\{\mathbf{W} \}_{i,j}= %
\cases{ 1, &\quad $\mbox{if States $
\mathbf{s}_i$ and $\mathbf{s}_j$ share a common border,}$
\vspace*{2pt}
\cr
0, &\quad $\mbox{otherwise}.$} %
\]

Then, since the general model described in Section \ref{secpatterns}
is overparameterized, it is necessary to impose some parameter
restrictions. For example, because $Y$ is univariate (i.e., $n_y=1$),
each column of $\mathbf{H}_y$ (i.e., $\mathbf{h}_{y_j}$) is treated as
a univariate GMRF with conditional mean
\[
\mathrm{E} \bigl[{h}_{y_j}(\mathbf{s}_i)|
\mathbf{R}_{-i} \bigr]= \mu _i^{({h}_{y_j})} +
\theta_{{h}_{y_j}} \sum_{u\in\mathcal{S}_i} \bigl({h}_{y_j}(
\mathbf{s}_u) - \mu_u^{({h}_{y_j})}\bigr)
\]
and conditional variance
\[
\operatorname{VAR} \bigl[{h}_{y_j}(\mathbf{s}_i)|
\mathbf{R}_{-i} \bigr]= \psi ^{({h}_{y_j})^2}.
\]

On the other hand, since $X$ is a bivariate process---that is, $n_x=2$
and $\mathbf{X}(\mathbf{s},t)= [{X}_1(\mathbf{s},t), {X}_2(\mathbf
{s},t) ]'$---we assume that for $i,u=1, \ldots,n$, $\mathbf
{T}_i^{(h_{x_j})}=\mathbf{T}^{(h_{x_j})}$ is a $(2 \times2)$
conditional covariance matrix and
\[
\mathbf{F}^{(h_{x_j})}=\mathbf{F}_{iu}^{(h_{x_j})}=-\left[
\matrix{ \theta_{x_1}^{(j)} &
\theta_{x_1,x_2}^{(j)}
\vspace*{2pt}\cr
\theta_{x_2,x_1}^{(j)} & \theta_{x_2}^{(j)}
}
\right].
\]

Hence, the covariance matrix can be written as
%
\[
\label{eqSigma1} \bolds{\Sigma}^{(h_{x_j})}= \bigl(\mathbf{I}_N
\otimes\mathbf {T}^{(h_{x_j})^{1/2}} \bigr) \bigl[\mathbf{I}_{\tilde{n}_x} + \mathbf
{W}^U \otimes\tilde{\mathbf{F}}^{(h_{x_j})}+\mathbf{W}^L
\otimes\tilde {\mathbf{F}}^{(h_{x_j})^\prime} \bigr]^{-1} \bigl(
\mathbf{I}_N \otimes \mathbf{T}^{(h_{x_j})^{1/2}} \bigr),
\]
where $\mathbf{W}^U$ and $\mathbf{W}^L$ denote the upper- and
lower-triangular parts of $\mathbf{W}$, respectively. Conditions for
which $\bolds{\Sigma}^{(h_{x_j})}$ is positive definite depend on the
parameter space of the spatial interaction parameters in $\mathbf
{F}^{(h_{{x_j}})}$. However, restricting $\bolds{\Sigma
}^{(h_{x_j})^{-1}}$ to be strictly diagonally dominant or adding a
penalty if some of the eigenvalues are negative will ensure positive
definitiveness [for a discussion on this point see \citet{SaiCre07}].

Since interpreting the spatial parameters in $\mathbf
{F}^{(h_{x_j})}$ requires some care, more information on the impact of
the choice of $\mathbf{F}^{(h_{x_j})}$ can be obtained by examining the
conditional covariance of two neighboring locations (given the rest)
\[
\boldSigma_{iu|-iu}^{h_{x_j}}=\left[ %
\matrix{\mathbf{T}_{i}^{(h_{x_j})^{-1}} & \mathbf{T}^{(h_{x_j})^{-1}}\mathbf
{F}^{(h_{x_j})}
\vspace*{2pt}\cr
\bigl(\mathbf{T}^{(h_{x_j})^{-1}}\mathbf{F}^{(h_{x_j})} \bigr)' &
\mathbf{T}^{(h_{x_j})^{-1}}
}
\right]^{-1}
\]
or, analogously, the conditional correlation matrix
%
\begin{equation}
\label{eqcondcorr} {\bolds{\Omega}}_{ij|-ij}={\boldDelta}^{-{1}/{2}}
\boldSigma _{iu|-iu}^{h_{x_j}}{\boldDelta}^{-{1}/{2}},
\end{equation}
where $\boldDelta=\operatorname{diag} (\boldSigma
_{iu|-iu}^{h_{x_j}} ).$\eject

The parameters for the priors on $\boldbeta^{(h_{x_i})}$,
$\mathbf{T}_i^{(h_{x_i})^{-1}}$ and $\tilde{\mathbf
{F}}_{iu}^{(h_{x_i})}$ are set as follows: $\sigma_\beta^2=100$,
$\varrho_x=20$, $\mathbf{S}_x=\mathbf{I}$ and $\varsigma=0.05$. The
design matrix $\bolds{\mathcal{D}}^*$ is specified to represent a constant
mean in space and we also consider $\mathbf{m}_y(t)=\mathbf{m}_y$ and
\mbox{$\mathbf{m}_x(t)=\mathbf{m}_x$}.

\textsc{Model specification}: \textit{State equation}.
Motivated by the debate on the possible existence of
cointegration between RHPI, RPCI and UR, we consider the cointegrated
model specification as shown in Section \ref{secsubnonstationary}.
The temporal lag of the state equations has been fixed to 2 (i.e.,
$p^*=2$), and an increasing number of common factors, that is, $2 \leq
m, l \leq12$, have been considered for the model specification. Then
the maximum possible number of cointegrating relationships is defined
as $r_d^*=m-1$ and $r_f^*=l-1$. Other modeling details, including prior
hyperparameter values, are defined in Section \ref{secinference} and
Appendix \ref{appSSVSprior}.

Together with the model specification described above, hereafter
denoted as M$_0$, other simpler models representing a simplification of
M$_0$ were also considered for comparison purposes. Specifically, to
have an idea of the relative importance of the different specifications
used in M$_0$ (e.g., correlated factor loadings and cointegrated
factors), three models with the following assumptions were considered:
(i) uncorrelated factor loadings and a simple VAR specification (i.e.,
without cointegration) for the state equation (M$_1$), (ii)
uncorrelated factor loadings and cointegrated factors (M$_2$), (iii)
correlated factor loadings and a simple VAR specification (i.e.,
without cointegration) for the factors (M$_3$). Finally, a fourth model
(M$_4$) which is relatively simple to estimate [see, e.g.,
Lutkepohl (\citeyear{Lut05})] but with a completely different structure is also considered:
\[
Y(\mathbf{s}_i,t)=\mathbf{c}(\mathbf{s}_i,t)'
\boldbeta(\mathbf{s}_i) + u_y(\mathbf{s}_i,t),
\]
where $\mathbf{c}(\mathbf{s}_i,t)$ is the vector containing
the covariates $X_1$ and $X_2$ (including the intercept), $\boldbeta
(\mathbf{s}_i)$ is the corresponding vector of (site-specific)
regression coefficients and $ u_y(\mathbf{s}_i,t)$ is a VAR$(2)$
process where the noise part of the model is assumed to be distributed
as a univariate GMRF (i.e., the noise is uncorrelated in time but it is
allowed to be spatially correlated). The introduction of a spatial
(GMRF) prior on the regression coefficients is also considered in the
parametrization.

\textsc{Model estimation}.
The identifiability constraints associated with the model to be
estimated concern the ordering of the States and the connection between
the chosen ordering and the specific form of the factor loading
matrices $\mathbf{H}_y$ and $\mathbf{H}_x$. Unfortunately, no fixed
rules exist to select the States which must be constrained. In the
following, we thus discuss a possible strategy which exploits results
from a cluster analysis performed (before estimating the model) on the
data matrices $\mathbf{Y}$ and $\mathbf{X}$, respectively, of
dimensions $(\tilde{n}_y \times T)$ and $(\tilde{n}_x \times T)$. In
this case, considering RHPI, the K-Means classification algorithm is
repetitively run for a number of clusters equal to $m$, with $ 2 \leq m
\leq12$. The States (one for each cluster) to be constrained are thus
chosen as the ones that: (possibly) belong to different BEA regions,
show the highest mean values of RHPI and/or are far apart from each
other (especially when $m$ is larger than the number of BEA regions).
For a given $l$, such that $2 \leq l \leq12$, the same procedure is
also applied to $\mathbf{X}$ and, whenever possible, the same States
selected for the housing prices are chosen. Note that especially in
cases in which $l>m$, the choice of the States within the clusters
obtained for $\mathbf{X}$ can be made independently of RHPI and based
on several criteria such as the membership to different BEA regions
and/or highest (smallest) mean values of RPCI (UR). When $m>l$, the
same criteria can be adopted to choose the States among the ones
already constrained in $\mathbf{H}_y$.

For each fitted model, the MCMC algorithm was run for $250\mbox{,}000$
iterations. Posterior inference was based on the last $150\mbox{,}000$ draws
using every $10$th member of the chain to avoid autocorrelation within
the sampled values. Several MCMC \textit{diagnostics} could be used to
test the convergence of the chains [see, e.g., \citet{Gew92},
Gilks, Richardson and Spiegelhalter (\citeyear{GilRicSpi96}), \citet{Spietal02}
and \citet{Jonetal06}].
In our case, convergence of the chains of the model was monitored
visually through trace plots as well as using the $R$-statistic of
\citet{Gel96} on four chains starting from very different values.

Competing models were compared using the predictive model choice
statistic, PMCC, described in Section \ref{secmodelselection}. The
$\mathrm{PMCC}$ criterion suggests that, for M$_0$, the optimal choice is
found with $m=7$ and $l=8$. The same number of components is also
confirmed for models M$_1$--M$_3$. However, compared with M$_3$, the
best of the three alternative models, the $\mathrm{PMCC}$ increases $17\%
$, which denotes much worse model fitting properties.

Notice that for M$_0$, the following States have been constrained in
the factor loading matrix $\mathbf{H}_y$: North Carolina, Montana,
California, Massachusetts, Texas, Illinois and Arizona. Instead,
considering $\mathbf{H}_x$, we have constrained $5$ States for UR:
North Carolina, California, Massachusetts, Texas and Illinois, and $3$
States for RPCI: Arizona, Montana and Massachusetts.

\begin{figure}

\includegraphics{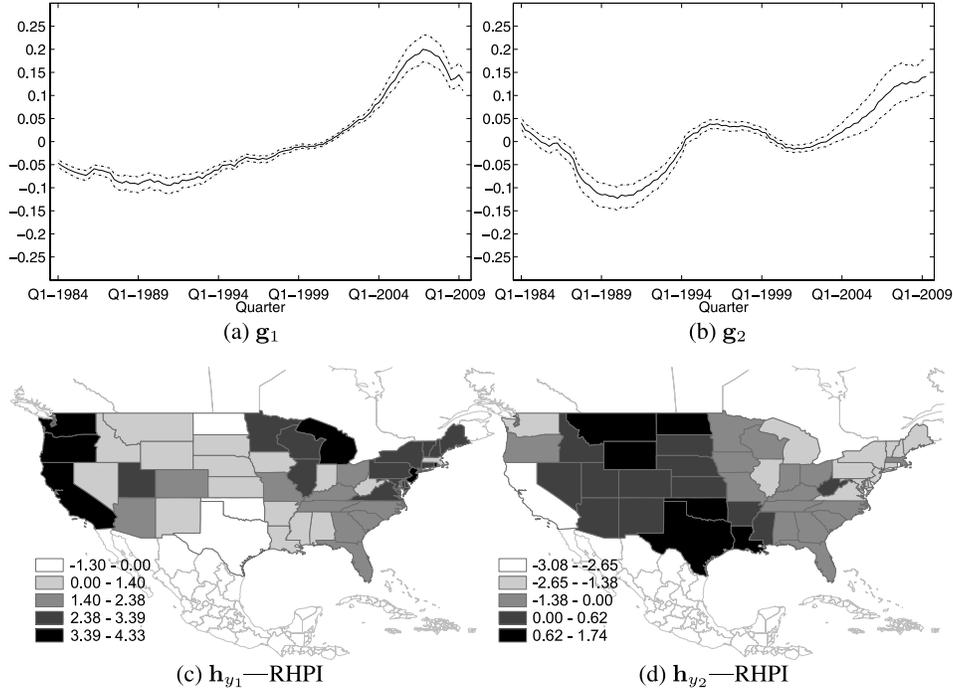}

\caption{Subplots \textup{(a)} and \textup{(b)}: marginal posterior medians for the
estimated latent factors $g_1(t)$ and $g_2(t)$ (continuous line) and
their $95 \%$ credible intervals (dashed line). Subplots \textup{(c)} and \textup{(d)}:
maps of the posterior medians for the factor loadings $\mathbf
{h}_{y_1}$ and $\mathbf{h}_{y_2}$ related to the real housing price
index.} \label{figg}
\end{figure}

\textsc{Factor loadings and common latent factors}.
The MCMC estimates of the endogenous components, $g_i(t),   i=1,\ldots,7$, appear as nonstationary processes, each representing
specific features of the large-scale temporal variability of the RHPI
series. The first two latent components represent common trends and are
characterized by narrow $95 \%$ credibility intervals. Specifically,
the pattern of the first component, shown in Figure~\ref{figg}(a),
highlights a growth of RHPI since the early nineties up to 2006
followed by a sustained decrease. At the national level, prices
increased substantially from 2000 to the peak in 2006 and then have
been falling very sharply across the country. An exploratory analysis
shows that this component tracks the pattern of the national RHPI,
although the latter seems to be a bit more volatile, especially in the
period 1984--1994. We also notice that this component is highly
correlated (i.e., the correlation is in general greater than $0.80$)
with all the State time series with the exception of Connecticut, Texas
and Oklahoma, for which the correlation is around $0.50$.

The series of the second component, $g_2(t)$, shown in Figure \ref
{figg}(b), is characterized by a price trough in the mid-1980s and
mid-1990s followed by a mild price peak. Then, the late 1990s begin
with a dramatic and sustained increase. Examination of the data plotted
in Figure \ref{figbea} shows that this is a typical pattern of the
$50\%$ of the States of Plains, Southeast and Rocky Mountain.

The remaining latent variables (not shown here) present some
peculiarities for the periods 1984--1990 and 2004--2007 and, compared
with the first two factors, are characterized by slightly wider
credibility intervals.

%

%
%

Figure \ref{figg}(c)--(d) show the maps of the estimated first two
factor loadings---that is, the first two columns of the measurement
matrix $\mathbf{H}_y$. The maps clearly show the presence of clusters
of US States. Table \ref{tabposteriorsummaryHy} also shows the
posterior summaries of the between-location conditional correlations
estimated [using equation (\ref{eqcondcorr})] for each column of $\mathbf
{H}_y$. Since the $95\%$ credibility intervals do not overlap zero and
all the conditional correlations seem to be statistically significant,
the clusters are easily identified by looking at the spatial patterns
of the factor loadings.

\begin{table}
\tabcolsep=0pt
\caption{Posterior summary of the between-location conditional
correlations for the columns of the measurement matrix $\mathbf{H}_y$.
In brackets we show the $2.5$ and $97.5$ percentiles used for defining
the $95\%$ credible interval limits}
\label{tabposteriorsummaryHy}
\begin{tabular*}{\textwidth}{@{\extracolsep{\fill}}lccccccc@{}}
\hline
& \multicolumn{7}{c@{}}{\textbf{Factor loadings} \textbf{(}$\mathbf{H_y}$\textbf{)}} \\[-4pt]
& \multicolumn{7}{c@{}}{\hrulefill} \\
& \textbf{1} & \textbf{2} & \textbf{3} & \textbf{4} & \textbf{5} & \textbf{6} & \textbf{7} \\
\hline
Median & 0.09 & 0.08 & 0.08 & 0.07 & 0.06 & 0.08 & 0.08 \\
$95\%$ CI & [0.05, 0.12] & [0.04, 0.12] & [0.02, 0.12] & [0.03, 0.10] &
[0.03, 0.12] & [0.02, 0.12] & [0.02, 0.12] \\
\hline
\end{tabular*}
\end{table}

Figure \ref{figg}(c) shows [using the \textit{natural break method} of
ArcMap, \citet{ESRI09}] the weights of the first factor loading, $\mathbf
{h}_{y_1}$. Except for Texas, Oklahoma and North Dakota, these weights
are all positive, with the highest loadings observed in the Pacific and
Northeast regions, which strongly influence the contiguous regions.

Figure \ref{figg}(d) also shows an interesting pattern in the
loadings. Southwest, Rocky Mountain States, some Plains States and
Louisiana have positive loadings, while the other States have negative
loadings. The States with highest loadings (Louisiana, New Mexico,
Texas, Oklahoma, North Dakota and Wyoming) show a temporal pattern very
similar to the second latent variables. On the other hand, the States
with lowest values (California, Connecticut, Michigan, New Jersey and
Rhode Island) show temporal dynamics which, at least until the end of
the nineties, result in the opposite of $g_2(t)$. Many of these States
in the last 25 years have been particular beneficiaries of new
technologies. These innovations interacting with restrictions on new
residential buildings have resulted in real housing prices in these
regions deviating from the average across US States over a relatively
prolonged period [Holly, Pesaran and
Yamagata (\citeyear{HolPesYam10})]. Also, considering the period
1984--1990, the spatial contrast highlighted in the map of Figure \ref
{figg}(d) clearly confirms that while West--South--Central regions
(especially ``oil-patch'' states such as Texas and Oklahoma) experienced
sharp declines, the Northeast and California housing market were
booming. Note that this map provides clear evidence of the results
described in Table \ref{tabcorr} where we have found significant
correlations between the States belonging to the East and West regions.

%
\begin{figure}

\includegraphics{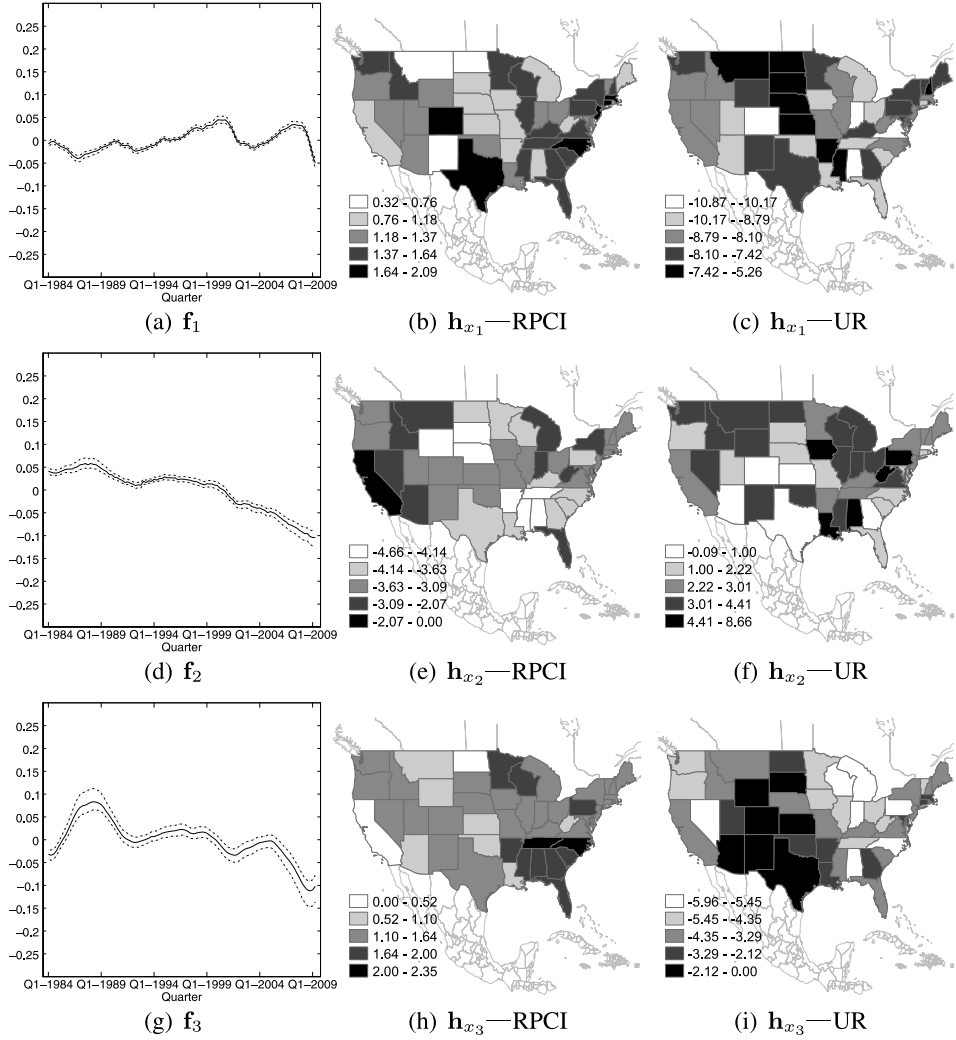}

\caption{Subplots \textup{(a)}, \textup{(d)} and \textup{(g)}: marginal posterior medians for the
estimated latent factors $f_1(t)$, $f_2(t)$ and $f_3(t)$ (continuous
line) and their $95 \%$ credible intervals (dashed line). Subplots \textup{(b)},
\textup{(e)} and \textup{(h)}: maps of the posterior medians for the factor loadings
$\mathbf{h}_{x_1}$, $\mathbf{h}_{x_2}$ and $\mathbf{h}_{x_3}$ related
to the real per capita personal income variable. Subplots \textup{(c)}, \textup{(f)} and
\textup{(i)}: maps of the posterior medians for the factor loadings $\mathbf
{h}_{x_1}$, $\mathbf{h}_{x_2}$ and $\mathbf{h}_{x_3}$ related to the
unemployment rate variable.} \label{figf}
\end{figure}

The MCMC estimates of the exogenous components, $f_i(t),   i=1,\ldots,8$, summarize the dynamics of RPCI and UR variables. The first three
of these latent factors, together with their $95 \%$ credibility
intervals, are shown in Figure \ref{figf}. These components seem to
have a substantial impact on RPCI and UR, although the latter shows
more complex dynamics which can be fully understood by examining the
behavior of all the estimated factors.

The first factor, ${f}_1(t)$, shows a cyclical behaviour with a
slightly positive trend in the period 1986--2000. The series exhibits a
trough in the period 2000--2006 followed by a sustained decrease. The
2000--2006 pattern has roots in the prior turmoil in the financial
markets. In fact, the period 2000--2001 is characterized by a rapid
decline of high tech industries, a collapse of the stock market and a
slow level of technology investment. The relaxed monetary policy
adopted by the Federal Reserve had thus lead to an increase of RPCI and
a decrease of UR up to~2007.

The factor loadings related to ${f}_1(t)$, shown in Figure \ref
{figf}(b) and Figure~\ref{figf}(c), are all positive for RPCI and
negative for UR. Figure \ref{figf}(b) clearly shows groups of States
with common spatial patterns. Specifically, we notice the presence of
two clusters: the first involves several States from the Great Lakes,
Southeast and New England, while the second is mainly characterized by
Oregon and some States of the Mountain region (Arizona, Utah, Nevada
and Wyoming). Also, the highest values are related to those States
(Colorado, Connecticut, Georgia, Massachusetts, New Jersey, North
Carolina and Texas) whose RPCI shows the same cyclical pattern of
${f}_1(t)$ in the period 1995--2009.

Figure \ref{figf}(c), related to UR, shows quite a big cluster of
States forming a ridge from Montana to Mississippi. For these States
the variations of UR are less pronounced with respect to those showing
the smallest loadings (e.g., Alabama, Colorado, Indiana and Virginia).

The dynamics of RPCI and UR in the first period of the series is
captured by the third latent factor ${f}_3(t)$ shown in Figure \ref
{figf}(g). The figure shows that the early nineties are characterized
by a trough of UR and a hill for the RPCI.

Figure \ref{figf}(h) shows a huge cluster with values of the loadings
in the range $1.10\mbox{--}1.64$; the highest values are observed in the
Southeast region for which the oscillations of RPCI are a bit more
pronounced than other States.

Figure \ref{figf}(i) shows that the States for which the trough of UR
is more pronounced are characterized by lowest values of the loadings.
Notice that this figure also shows a reasonable correspondence with
Figure \ref{figg}(d).

The second factor, ${f}_2(t)$, shows a decreasing trend associated with
negative values of $\mathbf{h}_{x_2}$---RPCI---and (mainly) positive
values of $\mathbf{h}_{x_2}$---UR. The maps of the factor loading
clearly provide information on those States which have experienced a
positive trend for RPCI (e.g., Alabama, Arkansas, Mississippi,
Nebraska, South Dakota, Tennessee and Wyoming) as well as a downward
trend for UR (see, e.g., Alabama, Iowa, Louisiana, Pennsylvania
and West Virginia).

%
%

The spatial structure of the factor loadings is also confirmed by the
the posterior summaries of their within- and between-location
conditional correlations and cross-correlations (see Table \ref
{tabposteriorsummaryHx}). The $95\%$ credibility intervals suggest
that most parts of these correlations can be considered as nonzero.
Also, the conditional spatial dependence of each factor loading is
positive, while
both the between- and the within-location conditional
cross-correlations are negative.

\begin{table}
\tabcolsep=0pt
\caption{Posterior summary of the within- and between-location
conditional correlations and cross-correlations for the first three
factor loadings columns related to the unemployment rate and real per
capita personal income variables. In brackets we show the $2.5$ and
$97.5$ percentiles used for defining the $95\%$ credible interval limits}
\label{tabposteriorsummaryHx}
\begin{tabular*}{\textwidth}{@{\extracolsep{\fill}}lccccc@{}}
\hline
& \multicolumn{5}{c@{}}{\textbf{Conditional correlation}} \\[-4pt]
& \multicolumn{5}{c@{}}{\hrulefill} \\
& \textbf{Within-location} & \textbf{Between-location} & \textbf{Between-location} &
\textbf{Between-location} & \textbf{Between-location} \\
& \textbf{RPCI vs UR} & \textbf{RPCI} & \textbf{RPCI vs UR} & \textbf{UR vs RPCI} & \textbf{UR}
\\
\hline
$\mathbf{h}_{x_1}$ & $-0.22$ & 0.06 & $-0.04$ & $-0.03$
& 0.05 \\
& [$-0.44$, $-0.09$] & [0.01, 0.09]
& [$-0.07$, $-0.02$] & [$-0.07$, $- 0.01$] & [0.03,
0.08] \\[3pt]
$\mathbf{h}_{x_2}$ & \phantom{$-$}0.02 & 0.05 & $-0.00$ & $-0.01$
& 0.07 \\
& [$-0.29$, 0.12] & [0.01, 0.10] &
[$-0.06$, 0.05] & [$-0.06$, 0.06] & [0.02, 0.09] \\[3pt]
$\mathbf{h}_{x_3}$ & $-0.27$ & 0.08 & $-0.02$ &
$-0.04$ & 0.07 \\
& [$-0.38$, $-0.02$] & [0.02, 0.12]
& [$-0.07$, 0.03] & [$-0.07$, $-0.01$] & [0.03, 0.10]
\\
\hline
\end{tabular*}
\end{table}

\textsc{Model estimation: Cointegration}.
As noted in the introduction, there has been quite a long debate in the
literature about whether there is cointegration between real
housing prices and fundamentals. The idea is that in the absence of
cointegration there are no fundamentals driving real housing prices and
the absence of an equilibrium relationship would essentially increase
the presence of bubbles [\citet{CasShi03}, Holly, Pesaran and
Yamagata (\citeyear{HolPesYam10})].
Here, we test the existence of this cointegrating relationship in a
latent space, avoiding to take account of the effect of the
cross-sectional dependence [see Holly, Pesaran and
Yamagata (\citeyear{HolPesYam10}) for a discussion on
this point]. In terms of cointegrated ranks, following \citet{Jocetal11},
our posteriors for $r_f$, $r_d$, $r_c$, $r_{c_1}$ and $r_{c_2}$
are obtained by considering the draws of their respective matrices
(i.e., $\boldPi_f$, $\boldPi_{gd}$, $\mathbf{A}\mathbf{B}_2^\prime
+\mathbf{A}_2\mathbf{B}_f^\prime$, $\mathbf{A}\mathbf{B}_2^\prime$ and
$\mathbf{A}_2\mathbf{B}_f^\prime$; see Appendix \ref{appcoint}) and
taking the number of singular values greater than $0.05$.

These are shown in Table \ref{tabposteriorrank} where we note that
there is a strong support for an exogenous cointegrated rank of either
4 or 5; for $r_d$ there is a hint of a rank equal to 5, but small
probabilities are also observed for 4 and 6. Finally, since there is
evidence that $r_c<r_{c_1}+r_{c_2}$, we may conclude that a
cointegration structure is confirmed between the endogenous and
exogenous processes. Such a result thus supports the idea about the
existence of a convergence to a stable equilibrium relationship and,
hence, about the absence of a US housing price bubble for the period
considered in the study.

\begin{table}
\caption{Posterior of cointegration ranks $r_f$, $r_d$, $r_c$,
$r_{c_1}$ and $r_{c_2}$}
\label{tabposteriorrank}
\begin{tabular*}{\textwidth}{@{\extracolsep{\fill}}lcccccc@{}}
\hline
& \multicolumn{6}{c@{}}{\textbf{Estimated probabilities for effective
ranks}}\\[-4pt]
& \multicolumn{6}{c@{}}{\hrulefill}\\
& \textbf{1} & \textbf{2} & \textbf{3} & \textbf{4} & \textbf{5} &\textbf{ 6} \\
\hline
$r_f$ & 0.00 & 0.00 & 0.01 & 0.34 & 0.61 & 0.04 \\
$r_d$ & 0.00 & 0.00 & 0.00 & 0.18 & 0.70 & 0.12 \\
$r_c$ & 0.00 & 0.00 & 0.00 & 0.02 & 0.48 & 0.50 \\
$r_{c_1}$ & 0.00 & 0.00 & 0.10 & 0.63 & 0.27 & 0.00 \\
$r_{c_2}$ & 0.00 & 0.00 & 0.03 & 0.45 & 0.50 & 0.02 \\
\hline
\end{tabular*}
\end{table}

To provide further evidence that our approach is yielding sensible
results, the use of Bayes factors using a non-SSVS prior
[\citet{Sug09},
\citet{KasRaf95}] confirms that, conditionally on $m=7$
and $l=8,$ results for $r_f$ and $r_d$ are similar to those presented here.

\textsc{Unconditional and conditional forecasts}.
To test the predictive performance of the SD-SEM model, the last 10
quarters have been excluded from the estimation procedure and used only
for forecast purposes. Hence, we consider the forecast for a horizon of
$k=10$ periods corresponding to the quarters $\mbox{Q3-2009--Q4-2011}$. Also, predictions of RHPI are obtained by following two settings:

\begin{longlist}[(ii)]
\item[(i)] \textit{unconditional predictions}: we only use past
information; hence, $X$ is not available for the forecast period;
\item[(ii)] \textit{conditional predictions}: the exogenous variables
$X_1$ and $X_2$ are assumed known in the period in which temporal
forecasts of RHPI are required.
\end{longlist}

\begin{figure}

\includegraphics{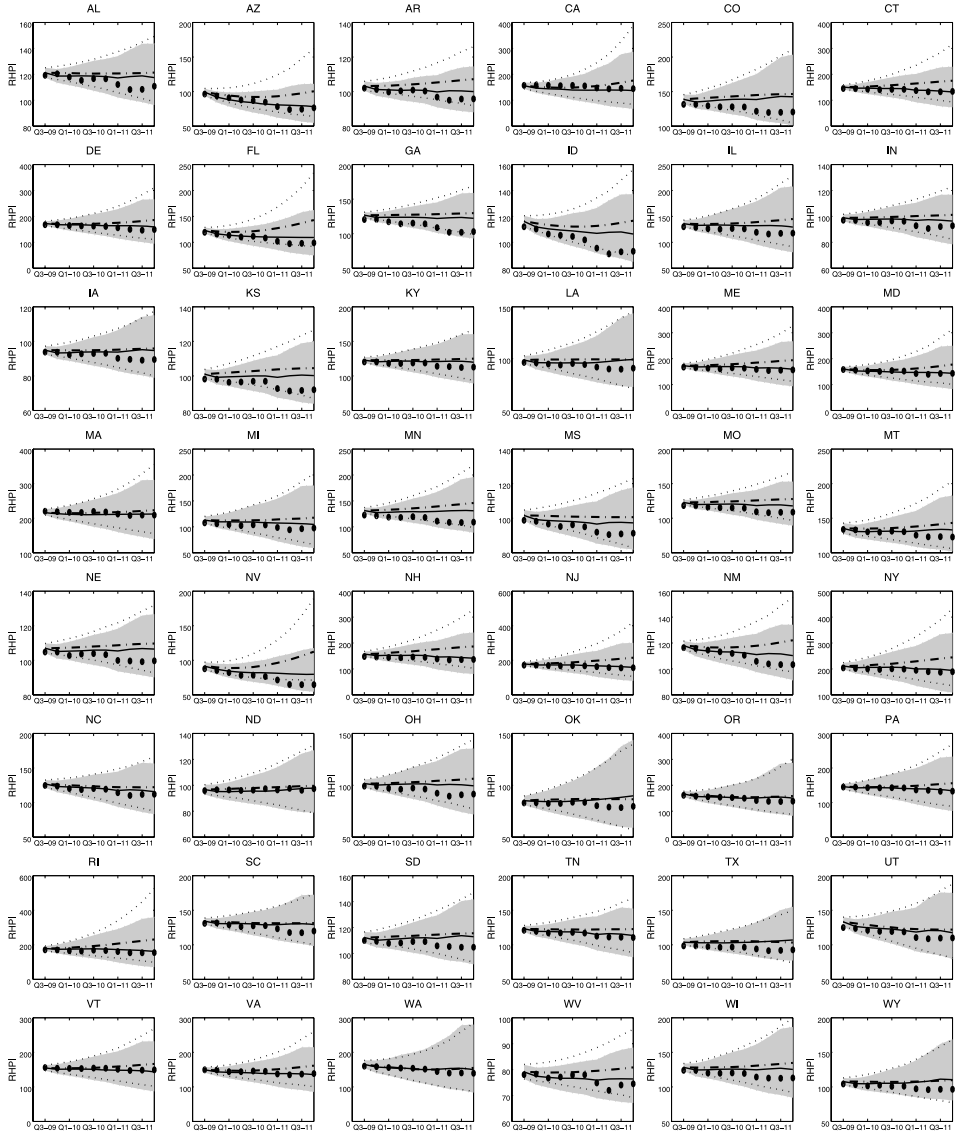}

\caption{Unconditional forecasts (dashed line), conditional forecasts
(continuous line) and true data ($\bullet$) at the 48 United States;
the $95\%$ credible interval limits for the unconditional forecasts are
represented by dotted lines. The $95\%$ credible interval limits for
the conditional forecasts are represented by the shaded area. Each
subplot also shows the initials of the State.}
\label{figforecasts1}
\end{figure}

For each State, both unconditional and conditional forecasts (together
with $95 \%$ credible intervals) of the housing price index are shown
in Figure \ref{figforecasts1}. In general, compared with true values,
good prediction results can be achieved and, as expected, the
conditional (on the known values of $X$) approach exhibits more
encouraging out-of-sample properties of the model, with data points
being more accurately predicted.

%

To provide some measures of goodness of prediction for the estimated
models, Table \ref{tableO3diagnosticprediction} gives details on the
root mean squared prediction error, $\mathrm{RMSE} =\sqrt{\operatorname{mean}  \{(\tilde{Y}(\mathbf{s},t)- E[\tilde{Y}(\mathbf
{s},t)_{\mathrm{rep}}])^2  \}}$, the mean absolute error deviation, $\mathrm
{MAE} = \operatorname{mean}  \{|\tilde{Y}(\mathbf{s},t)- E[\tilde
{Y}(\mathbf{s},t)_{\mathrm{rep}}]|  \}$ [where $\tilde{Y}$ is the variable
at the original scale and the mean is taken over the $({N \times k})$
observations], the coverage probabilities (CP) and the average width
(AIW) of the prediction intervals. We note that in the conditional case
model M$_0$ shows much smaller values for RMSE, MAE and AIW; on the
other hand, the coverage probabilities of the 95\% intervals are larger
than the nominal rate. Models M$_1$, M$_2$ and M$_3$ provide very
similar results and provide some hints on the role played by the
spatially autocorrelated factor loadings and cointegrated factors. In
general, model M$_0$ works better than M$_1$, M$_2$ and M$_3$ for which
the average width of the prediction intervals are wider. We note that
in introducing the spatial correlation the AIW reduces substantially.
The same effect, albeit with different intensity, can be observed
assuming cointegrated factors and this can be detected by contrasting
models M$_0$--M$_3$ and M$_1$--M$_2$.

By making the series stationary through a first difference
transformation, the best result of model M$_4$ is characterized by an
RMSE of $13.575$ and a MAE of $11.372$. This result is obtained by
using a GMRF prior on the regression coefficients. We also note that
for this model the regressors, $X_1$ and $X_2$, are assumed as known
for the forecast period. Producing unconditional predictions under
model M$_4$, in fact, is not straightforward since it requires further
adjustments for predicting the process $X$.

\begin{table}
\caption{Root mean squared prediction errors (RMSE), mean
absolute deviations (MAE), coverage probabilities (CP) and average
width (AIW) of the prediction intervals, for unconditional and
conditional forecasts of RHPI. The statistics are computed for the
estimated models \textup{M}$_0$, \textup{M}$_1$, \textup{M}$_2$, \textup{M}$_3$ and \textup{M}$_4$}
\label{tableO3diagnosticprediction}
\begin{tabular*}{\textwidth}{@{\extracolsep{\fill}}lcd{2.3}d{2.3}cd{3.3}@{}}
\hline
\textbf{Model} & \multicolumn{1}{c}{\textbf{Type of prediction}} & \multicolumn{1}{c}{\textbf{RMSE}} & \multicolumn{1}{c}{\textbf{MAE}} &
\multicolumn{1}{c}{\textbf{CP 95\% interval}} &
\multicolumn{1}{c@{}}{\textbf{AIW 95\% interval}} \\
\hline
M$_0$& Unconditional & 16.081 & 11.704 & 0.958 & 59.762 \\
& Conditional & 7.223 & 5.558 & 0.989 & 54.723 \\[3pt]
M$_1$& Unconditional & 17.294 & 12.950 & 1.000 & 140.052 \\
& Conditional & 9.497 & 6.614 & 1.000 & 138.140 \\[3pt]
M$_2$& Unconditional & 17.496 & 12.942 & 0.989 & 112.814 \\
& Conditional & 9.904 & 7.414 & 0.998 & 112.086 \\[3pt]
M$_3$& Unconditional & 17.150 & 12.759 & 0.969 & 77.042 \\
& Conditional & 9.331 & 6.695 & 0.985 & 76.445 \\[3pt]
M$_4$& Unconditional & \multicolumn{1}{c}{--} & \multicolumn{1}{c}{--} & \multicolumn{1}{c}{--} & \multicolumn{1}{c@{}}{--} \\
& Conditional & 13.575 & 11.372 & 0.920 & 53.180 \\
\hline
\end{tabular*}
\end{table}

\textsc{Multiplier analysis}.
We conclude the analysis by providing some results from multiplier
analysis [Lutkepohl (\citeyear{Lut05})] which is helpful to describe how the
housing price index reacts over time to exogenous impulses. In this
case, we can check if past values on either RPCI or UR, observed on a
specific State, contain useful information to predict the variation of
RHPI, in addition to the information on its past values.
It can be shown (see Appendix~\ref{appMult}) that the dynamic
multipliers, $\bolds{\Gamma}_k$, which reflect the marginal impacts of
changes in the predictors $X_1$ and $X_2$, are defined as
\[
\bolds{\Gamma}_k=\mathbf{H}_y\bolds{\mathfrak{J}} \bolds{
\mathcal{Q}}^k \bolds {\mathcal{B}}\mathbf{H}_x^\dag,\qquad
k=0, 1, \ldots,
\]
where, at the $k$th period (quarter), the $\gamma_{ij,k}$
element of the $(N \times\tilde{n}_x)$ matrix $\bolds{\Gamma}_k$
represents the response of the housing price in the $i$th State to a
given shock in the predictor $X_l,   l=1,2$, in State $j$, provided the
effect is not contaminated by other shocks to the system. The matrices
$\bolds{\mathfrak{J}}$, $\bolds{\mathcal{Q}}$ and $\bolds{\mathcal{B}}$, which
contribute to determine the multipliers, are defined in Appendix \ref{appMult}.

The impulse responses of RHPI to a $1\%$ shock in the exogenous
variables, RPCI and UR, in each State, show some interesting features.
However, since many possible interactions among States and variables
can be envisaged, in the following we provide a summary of the results
as well as a visual impression of some of the dynamic
interrelationships existing in the system. Note that following \citet{SimZha99} and
\citet{Pri05}, the credibility intervals of the
impulse response coefficients are discussed at the 16th and 84th
percentiles which, under normality, correspond to the bounds of a
one-standard-deviation.

One interesting feature is that a shock in RPCI in the States belonging
to New England (with the exception of Connecticut and New Jersey) does
not seem to produce evident effects on RHPI. The same holds for a RPCI
shock in Mideast States whose effects seem to disappear after one quarter.
It thus seems that past values of RPCI, in these regions, do not help
in forecasting RHPI throughout the US. At the same time, apart from New
Hampshire and Maryland, the prices in New England and the Mideast do
not seem to react to a RPCI shock in any other region. The housing
prices in Michigan, Ohio and Illinois, belonging to the Great Lakes,
also seem to behave similarly. Note that this similarity in behavior
was also found by \citet{ApePay12} in a study on housing price
convergence.

On the other hand, there is stronger evidence of the relationships
between UR shock effects in the States of New England and the Mideast
and RHPI responses in several States, mainly belonging to the
Southeast, Plains and Southwest regions. Also, RHPI forecasts in New
England and Mideast regions can be improved by exploiting UR
information on other States. In any case, considering the
infra-regional responses (i.e., RHPI responses of New England and
Mideast States to a UR shock produced in any State belonging to the
same region), we note that UR effects on the variation of RHPI
disappear after one period.

Regarding the remaining BEA regions, a $1\%$ shock to either RPCI or UR
seems to highlight effects on the housing prices involving quite a
large network of States, particularly in the second quarter. Analyzing
the impulse responses for longer periods, we note that the network of
relevant relationships between the States becomes sparser. However, the
most persistent effects on RHPI, which also involve a large numbers of
States belonging to the Southeast, Plains, Rocky Mountain, Southwest
and Far West regions, are associated to RPCI shocks in Nevada, Arizona,
Georgia, Alabama and Mississippi, and to UR shocks in Illinois, South
Carolina, Florida, Alabama, Iowa, South Dakota and Nebraska.

Moreover, the States whose RHPI responses are more persistent to RPCI
shocks in any other State of the aforementioned regions are Florida and
Nevada, while the States whose responses are more persistent to UR
shocks are New Mexico, Arizona, Arkansas and Mississippi.

If we consider the sign of the impulse response coefficients, we note
that, in general, a positive shock to RPCI is associated to a positive
effect on RHPI. Some exceptions are observed in the first period where
we can find negative coefficients. On the other hand, the scenario
appears to be different for the UR case, in which we note both positive
and negative effects on RHPI even for longer periods. Although we may
expect that unemployment has an adverse effect on real estate prices,
previous studies have nevertheless found unemployment to be positively
related to housing prices. For a discussion on this point we refer the
reader, for example, to \citet{VerVan09}, \citet{ClaMilPen10} and \citet{MoeNg11}.

\begin{figure}

\includegraphics{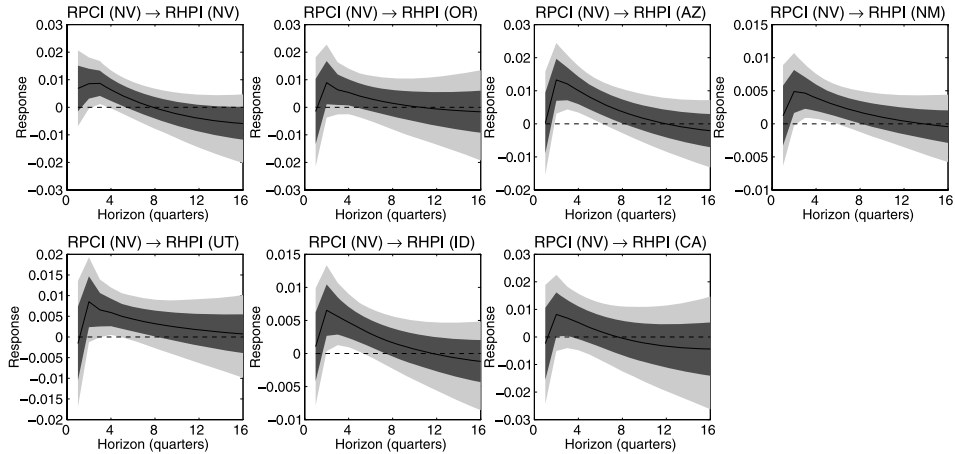}

\caption{Posterior mean impulse responses (solid line) of RHPI to a
RPCI shock in Nevada. The credibility intervals at $68\%$ and $90\%$
are represented by shaded areas. The responses are observed in Nevada
(NV), Oregon (OR), Arizona (AZ), New Mexico (NM), Utah (UT), Idaho (ID)
and California (CA).}\label{figimpatti1}
\end{figure}

\begin{figure}

\includegraphics{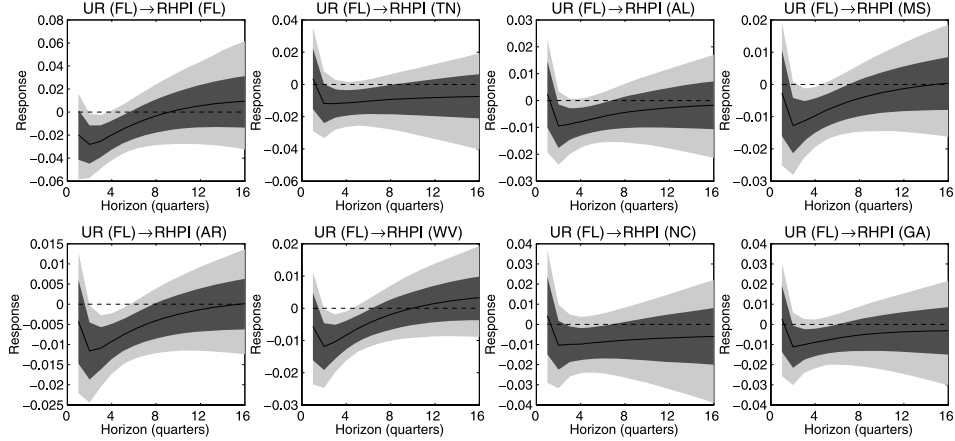}

\caption{Posterior mean impulse responses (solid line) of RHPI to a UR
shock in Florida. The credibility intervals at $68\%$ and $90\%$ are
represented by shaded areas. The responses are observed in Florida
(FL), Tennessee (TN), Alabama (AL), Mississippi (MS), Arkansas (AR),
West Virginia (WV), North Carolina (NC) and Georgia (GA).}
\label{figimpatti2}
\end{figure}

Finally, to provide a flavor of the type of relationship, Figure \ref
{figimpatti1} shows posterior mean housing price responses (solid
line) in Nevada, Oregon, Arizona, New Mexico, Utah, Idaho and
California to a $1\%$ shock to RPCI in Nevada. Figure \ref
{figimpatti2}, instead, shows the responses in Florida, Tennessee,
Alabama, Mississippi, Arkansas, West Virginia, North Carolina and
Georgia to a $1\%$ shock to UR in Florida. The shaded regions indicate
the credibility intervals corresponding to $68$ and $90$ percent.
Overall, the plots suggest that State-level responses follow a similar
pattern (consistently with the ripple effect) and, in most cases, the
effects tend to decay over two years, especially for UR shocks.

\section{Discussion}\label{secdiscussion}
In this paper we have discussed the modeling of spatio-temporal
multivariate processes observed on a lattice by means of
a Bayesian spatial dynamic structural equation model. We have used
ideas from factor analysis to frame and exploit both the spatial and the
temporal structure of the observed processes.

It can be shown that the SD-SEM encompasses a large class of
spatial-temporal models that are commonly used and,
more importantly, differs from them in two major aspects: (i) it avoids
the curse of dimensionality commonly present in
large spatio-temporal data and (ii) it facilitates the formation of
spatial clusters which further avoids dimensionality issues.

The model has been implemented in a Bayesian setup using MCMC sampling.
The MCMC chains of the parameters were monitored to detect possible
problems in convergence although no such problems were found in the
implementation.

The model was applied to study the impact that the real per capita
personal income and the unemployment rate may have on the real housing
prices in the USA using State level data.
Forecasting the future economic conditions and understanding the
relations between the observed variables have been two important
aspects covered by our model. The spatial variation is brought into the
model through the columns of the factor loading matrix and the
estimated conditional correlations and cross-correlations gave
significant evidence of spatial dependence associated with contiguity.
The spatial patterns of the factor loadings revealed several clusters
of interest showing common dynamics.

The time series dynamics have been captured by common dynamic factors.
An error correction model specification, with a cointegrating
relationship between the common latent factors, was found useful once
we took proper account of both heterogeneity and cross-sectional
dependence. Overall, results support the hypothesis that real housing
prices have been rising in line with fundamentals (real incomes and
unemployment rates), and there seems no evidence of housing price
bubbles at the national level.

Results from multiplier analysis were also helpful to describe how the
housing price index reacts over time to exogenous impulses. We have
found that, consistently with the ripple effect, the RHPI responses
show a similar pattern for neighboring States. The responses seem to be
more persistent to UR shocks, while the effects of a RPCI shock decay
more rapidly such that the system appears to approach faster to the
initial equilibrium conditions.

A further important advantage of the model formulation is that it
enables consideration of cases in which the temporal series of $X$ are
longer than those of $Y$. As noticed in Section \ref{secforecasting},
this was particularly useful to improve the temporal predictions by
conditioning on known values of the predictor providing a set of
plausible scenarios for RHPI.

Of course, we acknowledge that other possibilities could be considered
for modeling the spatial structure and an example is provided by \citet{WanWal03}. An alternative scheme could also lead to the
specification of common factors with a spatio-temporal structure. In
this case, one may follow the methodology proposed in \citet{DebErtLeS12} to quantify dynamic responses over time and space as
well as space--time diffusion impacts.

Finally, in this paper we have focused exclusively on normally
distributed data. However, nonlinear and non-Gaussian spatio-temporal
models have been extensively used in various areas of science, from
epidemiology to meteorology and environmental sciences, among others.
In this case, assuming the measurements belong to the exponential
family of distributions, a \textit{generalized} spatial dynamic
structural equation model represents a natural extension of the SD-SEM
discussed here. This extension will be a topic for future work.

\begin{appendix}\label{app}

\section{Cointegrated latent factors and their Vector Error Correction
representation}\label{appcoint}

Let $\tilde{\boldPhi}(z)$ denote the characteristic polynomial
associated with the vector ECM shown in (\ref{eqmodelnostationary})
and let $c$ be the number of unit roots of $\operatorname{Det} [\boldPhi(z)
]$. Let also that $\operatorname{rank}(\tilde{ \mathbf{A}}) = r$, with $r = m + l -
c$. Then, we assume that the latent exogenous variables, $\mathbf
{f}(t)$, are cointegrated with cointegrating rank $r_f$ so that $r >
r_f$ and $r_f < l$.

Let $\mathbf{Q} (\sum_{i=1}^{p} \boldPhi_i) \mathbf{P} =
\mathbf{J}$ be the Jordan canonical form of $\sum_{i=1}^{p} \boldPhi
_i$, where $\mathbf{Q}= \mathbf{P}^{-1}$ an $ ((m + l) \times(m +
l) )$ matrix, $\mathbf{J} = \operatorname{diag} (\mathbf{I}_{m - r_d},   \boldLambda_{r_d},   \mathbf{I}_{l-r_f},   \boldLambda_{r_f})$ and $r_d
\equiv r - r_f$ [\citet{AhnRei90} and \citet{Cho10}]. Because of the
exogeneity of $\mathbf{f}(t)$, the matrices $\tilde{\mathbf{A}}$ and
$\tilde{\boldPhi}_i$ are upper block triangular matrices, that is,
\[
\tilde{\mathbf{A}}= \left[ %
\matrix{\tilde{
\mathbf{A}}_1 & \tilde{\mathbf{A}}_{12}
\vspace*{2pt}\cr
\mathbf{0} & \tilde{\mathbf{A}}_2}
\right] \quad\mbox{and} \quad\tilde{\boldPhi}_i=
\left[ \matrix{ \tilde{\boldPhi}_{1i} & \tilde{
\boldPhi}_{12i}
\vspace*{2pt}\cr
\mathbf{0} & \tilde{\boldPhi}_{2i}}
\right].
\]
Then, consider the following matrix partition:
\[
\mathbf{P}=\left[ %
\matrix{ \mathbf{P}_1 &
\mathbf{P}_{12}
\vspace*{2pt}\cr
\mathbf{0} & \mathbf{P}_2}
\right],\qquad \mathbf{Q} = \mathbf{P} ^{-1}=\left[ \matrix{\mathbf{P}_1^{-1} & -
\mathbf{P}_{1}^{-1}\mathbf{P}_{12} \mathbf
{P}_{2}^{-1}
\vspace*{2pt}\cr
\mathbf{0} & \mathbf{P}_2^{-1}}
\right]=\left[\matrix{
\mathbf{Q}_1 & \mathbf{Q}_{12}
\vspace*{2pt}\cr
\mathbf{0} & \mathbf{Q}_2}
\right],
\]
with $\mathbf{Q}_1' = [{\mathbf{Q}_{1}^{(1)}\enskip     \mathbf
{Q}_{1}^{(2)}} ]$, $\mathbf{P}_1 = [{\mathbf{P}_{1}^{(1)} \enskip    \mathbf
{P}_{1}^{(2)}} ]$, $\mathbf{Q}_{12}' = [{\mathbf{Q}_{12}^{(1)}\enskip     \mathbf
{Q}_{12}^{(2)} }]$, $\mathbf{P}_{12} = [{\mathbf{P}_{12}^{(1)}  \enskip   \mathbf
{P}_{12}^{(2)}} ]$, $\mathbf{Q}_2' = [{\mathbf{Q}_{2}^{(1)}  \enskip   \mathbf
{Q}_{2}^{(2)} }]$ and $\mathbf{P}_2 = [{\mathbf{P}_{2}^{(1)}     \enskip\mathbf
{P}_{2}^{(2)}} ]$.

Note that $\mathbf{Q}_{1}^{(1)}$, $\mathbf{P}_{1}^{(1)}$ are
$ (m \times(m - r_d) )$, $\mathbf{Q}_{1}^{(2)}$, $\mathbf
{P}_{1}^{(2)}$ are $(m \times r_d)$, $\mathbf{Q}_{2}^{(1)}$, $\mathbf
{P}_{2}^{(1)}$ are $ (l \times(l-r_f) )$, $\mathbf
{Q}_{2}^{(2)}$, $\mathbf{P}_{2}^{(2)}$ are $(l \times r_f)$, $\mathbf
{Q}_{12}^{(1)}$ is $ (l \times(m-r_d) )$, $\mathbf
{P}_{12}^{(1)}$ is $ (m \times(l-r_f) )$, $\mathbf
{Q}_{12}^{(2)}$ is $(l \times r_d)$ and $\mathbf{P}_{12}^{(2)}$ is $(m
\times r_f).$ Then, we may write
%
\begin{eqnarray*}
\tilde{\mathbf{A}} & = & -\mathbf{P}(\mathbf{I} -\mathbf{J}) \mathbf {Q}= -
\left[ \matrix{ \mathbf{P}_{1}^{(2)}
& \mathbf{P}_{12}^{(2)}
\vspace*{2pt}\cr
\mathbf{0} & \mathbf{P}_{2}^{(2)}}
\right] \left[ \matrix{
\mathbf{I}- \boldLambda_{r_d} & \mathbf{0}
\vspace*{2pt}\cr
\mathbf{0} & \mathbf{I}- \boldLambda_{r_f}}
\right] \left[ \matrix{
\mathbf{Q}_{1}^{(2)'} & \mathbf{Q}_{12}^{(2)'}
\vspace*{2pt}\cr
\mathbf{0} & \mathbf{Q}_{2}^{(2)'}}
\right]
\\
& = & -\left[ %
\matrix{\mathbf{P}_{1}^{(2)}
(\mathbf{I}- \boldLambda_{r_d}) \mathbf {Q}_{1}^{(2)'}
& \mathbf{P}_{1}^{(2)} (\mathbf{I}- \boldLambda_{r_d})
\mathbf{Q}_{1}^{(2)'}\mathbf{P}_{12}
\mathbf{Q}_{2}+ \mathbf {P}_{12}^{(2)} (\mathbf{I}-
\boldLambda_{r_f}) \mathbf{Q}_{2}^{(2)'}
\vspace*{2pt}\cr
\mathbf{0} & \mathbf{P}_{2}^{(2)} (\mathbf{I}-
\boldLambda_{r_f}) \mathbf{Q}_{2}^{(2)'}}
\right],
\end{eqnarray*}
and equation (\ref{eqmodelnostationary}) can thus be
rewritten as
%
\begin{eqnarray}
\Delta\mathbf{g}(t) &= & \mathbf{A}\mathbf{B}'\mathbf{d}(t-1) +
\mathbf {A}_{2}\mathbf{B}_f'\mathbf{f}(t-1)+
\sum_{i=1}^{p-1}\mathbf{K}_i
\Delta\mathbf{d}(t-i)+\boldxi(t),\label{eqstatedg}
\\
\Delta\mathbf{f}(t) &= & \mathbf{A}_f\mathbf{B}_f'
\mathbf{f}(t-1) + \sum_{i=1}^{p-1}\tilde{
\boldPhi}_{2i} \Delta\mathbf{f}(t-j)+\boldeta (t),\label{eqstatef}
\end{eqnarray}
where $\mathbf{A} = -\mathbf{P}_{1}^{(2)} (\mathbf{I}-
\boldLambda_{r_d})$, $\mathbf{B} = [{\mathbf{I}\enskip      -\mathbf
{P}_{12}\mathbf{Q}_{2}}]'\mathbf{Q}_{1}^{(2)}$, $\mathbf{A}_f = -\mathbf
{P}_{2}^{(2)} (\mathbf{I}- \boldLambda_{r_f}) $, $\mathbf{A}_{2} =
-\mathbf{P}_{12}^{(2)} (\mathbf{I}- \boldLambda_{r_f}) $, $\mathbf{B}_f
= \mathbf{Q}_{2}^{(2)}$ and $\mathbf{K}_i= [{\tilde{\boldPhi}_{1i} \enskip    \tilde{\boldPhi}_{12i}} ] $. Note that if $\mathbf{P}_{12}$ and $\mathbf
{P}_{12}^{(2)}$ are $\mathbf{0}$, then a separated cointegrated
structure exists for $\mathbf{g}(t)$ and $\mathbf{f}(t)$.

Let $\mathbf{B} = [{\mathbf{B}_1^\prime \enskip   \mathbf{B}_2^\prime}
]^\prime$ where $\mathbf{B}_1= \mathbf{Q}_{1}^{(2)}$ and $\mathbf{B}_2=
-\mathbf{Q}_{2}' \mathbf{P}_{12}'\mathbf{Q}_{1}^{(2)}$, then $\tilde
{\mathbf{A}}$ can be rewritten as
%
\[
\tilde{\mathbf{A}}  =  -\left[\matrix{  \mathbf{A}
\mathbf{B}_1^\prime& \mathbf{A}\mathbf{B}_2^\prime+
\mathbf {A}_2\mathbf{B}_f^\prime
\vspace*{2pt}\cr
\mathbf{0} & \mathbf{A}_f\mathbf{B}_f^\prime}
\right].
\]

Also, let $r_f=\operatorname{rank}(\mathbf{A}_f\mathbf{B}_f^\prime)$,
$r_d=\operatorname{rank}(\mathbf{A}\mathbf{B}^\prime)$, $r_c=\operatorname{rank}(\mathbf{A}\mathbf
{B}_2^\prime+\mathbf{A}_2\mathbf{B}_f^\prime)$, $r_{c_1}=\operatorname{rank}(\mathbf
{A}\mathbf{B}_2^\prime)$ and $r_{c_2}=\operatorname{rank}(\mathbf{A}_2\mathbf
{B}_f^\prime).$ Then, it follows that if $\operatorname{rank}(\mathbf{A}\mathbf
{B}_2^\prime+\mathbf{A}_2\mathbf{B}_f^\prime)=0$, no cointegration
structure exists between the endogenous and exogenous processes $\mathbf
{g}(t)$ and $\mathbf{f}(t)$.

\section{The SSVS prior for the vector ECM}\label{appSSVSprior}

Since $\boldPi_{gd} = \mathbf{A}\mathbf{B}'$, $\boldPi_{gf} =
\mathbf{A}_{2}\mathbf{B}_f'$ and $\boldPi_{f}=\mathbf{A}_f\mathbf
{B}_f'$ are not unique, in this paper we follow the approach proposed
by \citet{Jocetal11} and Koop, Le{\'o}n-Gonz{\'a}lez and
Strachan (\citeyear{KooLeoStr10}) to elicit the SSVS
priors on the cointegration space. A summary of this approach is
provided below.

Specifically, a nonidentified $r^*_d \times r^*_d$ symmetric
positive definite matrix $\mathbf{E}$ is introduced with the property,
$\boldPi_{gd} = \mathbf{A}\mathbf{E} \mathbf{E}^{-1}\mathbf{B}'\equiv
\bar{\mathbf{A}}\bar{\mathbf{B}}'$, where $\bar{\mathbf{A}}=\mathbf
{A}\mathbf{E}$ and $\bar{\mathbf{B}}=\mathbf{B}\mathbf{E}^{-1}$. The
introduction of the nonidentified matrix $\mathbf{E}$ facilitates
posterior computation because the posterior conditional distributions
of $\bar{\mathbf{A}}$ and $\bar{\mathbf{B}}$ in the MCMC algorithm are
Gaussian [Koop, Le{\'o}n-Gonz{\'a}lez and
Strachan (\citeyear{KooLeoStr10})]. The same holds analogously for $\boldPi
_{gf}$ and $\boldPi_{f}$.

Let $\bar{\mathbf{a}}=\operatorname{vec}(\bar{\mathbf{A}}')$ and $\boldrho
=(\rho_1, \ldots, \rho_{\tilde{m}})$ a parameter vector, where $\tilde
{m}=m r^*_d$. Then, we assume that $\bar{\mathbf{a}}| \boldrho\sim
N(\mathbf{0}, \mathbf{V}_{0})$, where $\mathbf{V}_{0}=\operatorname{diag}(v_1^2,\ldots,
v_{\tilde{m}}^2)$, $v_{i}^2= ( 1- \rho_i) v_{0i}^2 + \rho_i v_{1i}^2$
and $\rho_i$, the $i$th element of $\boldrho$, has a Bernoulli
distribution with parameter $p_a$, that is, $\rho_i \sim \operatorname{Be}(p_a)$. In
this paper, we set $p_a=0.5$, $v_{0i}^2= 0.1 \hat{\sigma}^2(\bar
{a}_i)$, $v_{1i}^2= 10 \hat{\sigma}^2(\bar{a}_i)$, where $\hat{\sigma
}^2(\bar{a}_i)$ is an estimate of the variance of the $i$th element of
$\bar{\mathbf{a}}$ obtained from a preliminary MCMC run with a
noninformative prior.

With appropriate notation, the same assumptions hold for $\bar
{\mathbf{a}}_f =\operatorname{vec} (\bar{\mathbf{A}}_f)$, with $\bar{\mathbf{A}}_f =
\mathbf{A}_f \mathbf{E}_f$, and $\bar{\mathbf{a}}_2 =\operatorname{vec} (\bar{\mathbf
{A}}_2)$, with $\bar{\mathbf{A}}_2 = \mathbf{A}_2 \mathbf{E}_f$.

The prior for the cointegrated space is defined through $\bar
{\mathbf{b}} \sim N(\mathbf{0}, \mathbf{I})$ and $\bar{\mathbf{b}}_f
\sim N(\mathbf{0}, \mathbf{I})$, where $\bar{\mathbf{b}}=\operatorname{vec}(\bar
{\mathbf{B}})$, $\bar{\mathbf{b}}_f=\operatorname{vec}(\bar{\mathbf{B}}_f)$ and $\bar
{\mathbf{B}}_f=\mathbf{B}\mathbf{E}_f^{-1}$. The SSVS prior for $\mathbf
{k}=\operatorname{vec}([\mathbf{K}_1, \ldots, \mathbf{K}_{p^*-1}]')$ is given by
$\mathbf{k}|\bolddelta\sim N(\mathbf{0}, \mathbf{D})$, where $\mathbf
{D}=\operatorname{diag}  (\tau_1^2,\ldots,\tau_{(m+l)(p^*-1)}^2  )$, $\tau
_i^2=(1-\delta_i)\tau_{0i}^2 + \delta_i\tau_{1i}^2$ and $\bolddelta$ is
an unknown vector with typical element $\delta_i \sim \operatorname{Be}(p_{\tau})$.
Here, we set $p_\tau=0.5$, $\tau_{0i}^2= 0.1 \hat\sigma^2(k_i)$, $\tau
_{1i}^2= 10 \hat\sigma^2(k_i)$, and $\hat\sigma^2(k_i)$ is an estimate
of the variance of the $i$th element of $\mathbf{k}$ obtained from a
preliminary MCMC run using a noninformative prior. Analogously, we
define $\boldphi=\operatorname{vec}([\tilde{\boldPhi}_{21}, \ldots, \tilde{\boldPhi
}_{2p^*-1}]')$ and assume that $\boldphi|\bolddelta_\phi\sim N(\mathbf
{0}, \mathbf{D}_\phi)$, where $\mathbf{D}_\phi=\operatorname{diag}  (\kappa_{\phi
1}^{2},\ldots,\kappa_{\phi m^2(p^*-1)}^{2}  )$, $\kappa_{\phi
i}^{2}=(1-\delta_{\phi i})\kappa_{\phi0i}^{2} + \delta_{\phi i}\kappa
_{\phi1i}^{2}$ and $\bolddelta_\phi$ is an unknown vector with element
$\delta_{\phi i} \sim \operatorname{Be}(p_{\phi})$. Here we set $p_\phi=0.5$, $\kappa
_{\phi0i}^{2}= 0.1 \hat\sigma^2(\bar{\phi}_i)$, $\kappa_{\phi1i}^{2}=
10 \hat\sigma^2(\bar{\phi}_i)$, and $\hat\sigma^2(\bar{\phi}_i)$ is an
estimate of the variance of the $i$th element of $\bar{\bolds{\phi}}$
obtained from a preliminary MCMC run using a noninformative prior.

\section{Multiplier analysis}\label{appMult}

If the model contains integrated variables and the generation mechanism
is started at time $t = 0$, it readily follows that [L\"{u}tkepohl (\citeyear{Lut05}),
page 402--407]
%
\begin{equation}
\label{eqappc} \mathbf{g}(t) = \bolds{\mathfrak{J}}\bolds{\mathcal{Q}}^t
\mathbf{g}(0) + \sum_{i=0}^{t-1}\bolds{
\mathfrak{J}} \bolds{\mathcal{Q}}^i \bolds{\mathcal {B}}\mathbf{f}(t-i)+
\sum_{i=0}^{t-1}\bolds{\mathfrak{J}} \bolds{
\mathcal {Q}}^i \bolds{\mathfrak{J}}'\boldxi(t-i),
\end{equation}
where $\bolds{\mathfrak{J}}$, $\bolds{\mathcal{B}}$ and $\bolds
{\mathcal{Q}}$ are $ (m \times(m p + l s) )$, $ ((mp+ ls)
\times l  )$ and $ ((mp+ ls) \times(mp+ ls) )$ matrices
such that
\begin{eqnarray*}
\bolds{\mathfrak{J}}&=&\left[\matrix{ \mathbf{I} &
\mathbf{0} & \cdots& \mathbf{0} }
\right], \\
\bolds{\mathcal{B}}&=&\left[\matrix{\mathbf{0}
\vspace*{2pt}\cr
\mathbf{0}
\vspace*{2pt}\cr
\vdots
\vspace*{2pt}\cr
\mathbf{0}
\vspace*{2pt}\cr
\mathbf{I}_l
\vspace*{2pt}\cr
\mathbf{0}
\vspace*{2pt}\cr
\vdots
\vspace*{2pt}\cr
\mathbf{0}}\right],\qquad \bolds{\mathcal{Q}}=\left[ \matrix{ \mathbf{C}_1 & \mathbf{C}_2 &
\cdots& \mathbf{C}_p & \mathbf{D}_1 & \cdots&
\mathbf{D}_2 & \mathbf{D}_s
\vspace*{2pt}\cr
\mathbf{I}_m & \mathbf{0} & \cdots& \mathbf{0} & \mathbf{0} &
\cdots & \mathbf{0} & \mathbf{0}
\vspace*{2pt}\cr
\vdots& \vdots & \vdots& \vdots& \vdots& \vdots & \vdots& \vdots
\vspace*{2pt}\cr
\mathbf{0} & \cdots & \mathbf{I}_m & \mathbf{0} & \mathbf{0} &
\cdots & \mathbf{0} & \mathbf{0}
\vspace*{2pt}\cr
\mathbf{0} & \cdots& \mathbf{0} & \mathbf{0} & \mathbf{0} & \mathbf {0} &
\cdots& \mathbf{0}
\vspace*{2pt}\cr
\mathbf{0} & \cdots& \mathbf{0} & \mathbf{0} & \mathbf{I}_l &
\mathbf {0} & \cdots& \mathbf{0}
\vspace*{2pt}\cr
\vdots& \vdots & \vdots& \vdots& \vdots& \ddots & \vdots& \vdots
\vspace*{2pt}\cr
\mathbf{0} & \cdots& \mathbf{0} & \mathbf{0} & \mathbf{0} & \cdots&
\mathbf{I}_l & \mathbf{0}}
\right].
\end{eqnarray*}

Then, assuming without loss of generality $\mathbf
{m}_x(t)=\mathbf{0}$ and $\mathbf{m}_y(t)=\mathbf{0}$, it follows from
the measurement equation (\ref{eqmeasx}) that by denoting with
$\mathbf{H}_x^\dag$ the pseudo-inverse of $\mathbf{H}_x$, that is,
$\mathbf{H}_x^\dag=(\mathbf{H}_x'\mathbf{H}_x)^{-1}\mathbf{H}_x'$, for
$m<\tilde{n}_x$ and $\mathbf{H}_x'\mathbf{H}_x$ invertible, the
least-square estimator of $\mathbf{f}(t)$ is $\hat{\mathbf
{f}}(t)=\mathbf{H}_x^\dag\mathbf{X}(t).$

Hence, from equations (\ref{eqmeasy}) and (\ref{eqappc}),
it follows that the marginal impact of changes of the predictor $\mathbf
{X}(t)$ on the dependent variable $\mathbf{Y}(t)$ can be investigated
through the coefficient matrices
\[
\bolds{\Gamma}_k=\mathbf{H}_y\bolds{\mathfrak{J}} \bolds{
\mathcal{Q}}^k \bolds {\mathcal{B}}\mathbf{H}_x^\dag,\qquad
k=0, 1, \ldots
\]
\end{appendix}
\section*{Acknowledgments}
The authors would like to thank the Editor, the Associate
Editor and the two anonymous referees for helpful comments and
suggestions which have significantly improved the quality of the paper.
The authors are also very grateful to G. Koop and G.~J. Holloway for
invaluable comments on preliminary versions.


%


\printaddresses


\begin{thebibliography}{65}

\bibitem[\protect\citeauthoryear{Ahn and Reinsel}{1990}]{AhnRei90}
\begin{barticle}[mr]
\bauthor{\bsnm{Ahn},~\bfnm{Sung~K.}\binits{S.~K.}} \AND
\bauthor{\bsnm{Reinsel},~\bfnm{Gregory~C.}\binits{G.~C.}}
(\byear{1990}).
\btitle{Estimation for partially nonstationary multivariate autoregressive
models}.
\bjournal{J. Amer. Statist. Assoc.}
\bvolume{85}
\bpages{813--823}.
\bid{issn={0162-1459}, mr={1138362}}
\bptok{imsref}%
\end{barticle}
\endbibitem

\bibitem[\protect\citeauthoryear{Anselin}{1988}]{Ans88}
\begin{bbook}[auto:STB|2013/02/26|09:05:15]
\bauthor{\bsnm{Anselin},~\bfnm{L.}\binits{L.}}
(\byear{1988}).
\btitle{Spatial Econometrics: Models and Applications}.
\bpublisher{Kluwer Academic}, \blocation{Dordrecht, The Netherlands}.
\bptok{imsref}%
\end{bbook}
\endbibitem

\bibitem[\protect\citeauthoryear{Apergis and Payne}{2012}]{ApePay12}
\begin{barticle}[auto:STB|2013/02/26|09:05:15]
\bauthor{\bsnm{Apergis},~\bfnm{N.}\binits{N.}} \AND
\bauthor{\bsnm{Payne},~\bfnm{J.~E.}\binits{J.~E.}}
(\byear{2012}).
\btitle{Convergence in U.S. housing prices by state: Evidence from the club
convergence and clustering procedure}.
\bjournal{Letters in Spatial and Resource Sciences}
\bvolume{5}
\bpages{103--111}.
\bptok{imsref}%
\end{barticle}
\endbibitem

\bibitem[\protect\citeauthoryear{Banerjee, Carlin and
Gelfand}{2004}]{BanCarGel04}
\begin{bbook}[auto:STB|2013/02/26|09:05:15]
\bauthor{\bsnm{Banerjee},~\bfnm{S.}\binits{S.}},
\bauthor{\bsnm{Carlin},~\bfnm{B.~P.}\binits{B.~P.}} \AND
\bauthor{\bsnm{Gelfand},~\bfnm{A.~E.}\binits{A.~E.}}
(\byear{2004}).
\btitle{Hierarchical Modeling and Analysis for Spatial Data}.
\bpublisher{Chapman \& Hall/CRC}, \blocation{Boca Raton, FL}.
\bptok{imsref}%
\end{bbook}
\endbibitem

\bibitem[\protect\citeauthoryear{Box, Jenkins and Reinsel}{1994}]{BoxJenRei94}
\begin{bbook}[mr]
\bauthor{\bsnm{Box},~\bfnm{George E.~P.}\binits{G.~E.~P.}},
\bauthor{\bsnm{Jenkins},~\bfnm{Gwilym~M.}\binits{G.~M.}} \AND
\bauthor{\bsnm{Reinsel},~\bfnm{Gregory~C.}\binits{G.~C.}}
(\byear{1994}).
\btitle{Time Series Analysis: Forecasting and Control},
\bedition{3rd} ed.
\bpublisher{Prentice Hall Inc.}, \blocation{Englewood Cliffs, NJ}.
\bid{mr={1312604}}
\bptok{imsref}%
\end{bbook}
\endbibitem

\bibitem[\protect\citeauthoryear{Brown, Vannucci and Fearn}{1998}]{BroVanFea98}
\begin{barticle}[mr]
\bauthor{\bsnm{Brown},~\bfnm{P.~J.}\binits{P.~J.}},
\bauthor{\bsnm{Vannucci},~\bfnm{M.}\binits{M.}} \AND
\bauthor{\bsnm{Fearn},~\bfnm{T.}\binits{T.}}
(\byear{1998}).
\btitle{Multivariate {B}ayesian variable selection and prediction}.
\bjournal{J. R. Stat. Soc. Ser. B Stat. Methodol.}
\bvolume{60}
\bpages{627--641}.
\bid{doi={10.1111/1467-9868.00144}, issn={1369-7412}, mr={1626005}}
\bptok{imsref}%
\end{barticle}
\endbibitem

\bibitem[\protect\citeauthoryear{Cameron, Muellbauer and
Murphy}{2006}]{CamMueMur06}
\begin{bbook}[auto:STB|2013/02/26|09:05:15]
\bauthor{\bsnm{Cameron},~\bfnm{G.}\binits{G.}},
\bauthor{\bsnm{Muellbauer},~\bfnm{J.}\binits{J.}} \AND
\bauthor{\bsnm{Murphy},~\bfnm{A.}\binits{A.}}
(\byear{2006}).
\btitle{Was There a British House Price Bubble? Evidence from a Regional Panel.
Mimeo}.
\bpublisher{Oxford Univ. Press}, \blocation{London}.
\bptok{imsref}%
\end{bbook}
\endbibitem

\bibitem[\protect\citeauthoryear{Capozza et~al.}{2002}]{Capetal02}
\begin{bmisc}[auto:STB|2013/02/26|09:05:15]
\bauthor{\bsnm{Capozza},~\bfnm{D.~R.}\binits{D.~R.}},
\bauthor{\bsnm{Hendershott},~\bfnm{P.~H.}\binits{P.~H.}},
\bauthor{\bsnm{Mack},~\bfnm{C.}\binits{C.}} \AND
\bauthor{\bsnm{Mayer},~\bfnm{C.~J.}\binits{C.~J.}}
(\byear{2002}).
\bhowpublished{Determinants of real house price dynamics. NBER Working Paper 9262.}
\bptok{imsref}%
\end{bmisc}
\endbibitem

\bibitem[\protect\citeauthoryear{Carter and Kohn}{1994}]{CarKoh94}
\begin{barticle}[mr]
\bauthor{\bsnm{Carter},~\bfnm{C.~K.}\binits{C.~K.}} \AND
\bauthor{\bsnm{Kohn},~\bfnm{R.}\binits{R.}}
(\byear{1994}).
\btitle{On {G}ibbs sampling for state space models}.
\bjournal{Biometrika}
\bvolume{81}
\bpages{541--553}.
\bid{doi={10.1093/biomet/81.3.541}, issn={0006-3444}, mr={1311096}}
\bptok{imsref}%
\end{barticle}
\endbibitem

\bibitem[\protect\citeauthoryear{Case and Shiller}{2003}]{CasShi03}
\begin{barticle}[auto:STB|2013/02/26|09:05:15]
\bauthor{\bsnm{Case},~\bfnm{K.~E.}\binits{K.~E.}} \AND
\bauthor{\bsnm{Shiller},~\bfnm{R.~J.}\binits{R.~J.}}
(\byear{2003}).
\btitle{Is there a bubble in the housing market?}
\bjournal{Brookings Papers on Economic Activity}
\bvolume{2}
\bpages{299--362}.
\bptok{imsref}%
\end{barticle}
\endbibitem

\bibitem[\protect\citeauthoryear{Cho}{2010}]{Cho10}
\begin{bmisc}[auto:STB|2013/02/26|09:05:15]
\bauthor{\bsnm{Cho},~\bfnm{S.}\binits{S.}}
(\byear{2010}).
\bhowpublished{Inference of cointegrated model with exogenous variables. SIRFE
Working Paper~10--A04.}
\bptok{imsref}%
\end{bmisc}
\endbibitem

\bibitem[\protect\citeauthoryear{Clayton, Miller and Peng}{2010}]{ClaMilPen10}
\begin{barticle}[auto:STB|2013/02/26|09:05:15]
\bauthor{\bsnm{Clayton},~\bfnm{J.}\binits{J.}},
\bauthor{\bsnm{Miller},~\bfnm{N.}\binits{N.}} \AND
\bauthor{\bsnm{Peng},~\bfnm{L.}\binits{L.}}
(\byear{2010}).
\btitle{Price-volume correlation in the housing market: Causality and
co-movements}.
\bjournal{Journal of Real Estate Finance and Economics}
\bvolume{40}
\bpages{14--40}.
\bptok{imsref}%
\end{barticle}
\endbibitem

\bibitem[\protect\citeauthoryear{Cressie}{1993}]{Cre93}
\begin{bbook}[mr]
\bauthor{\bsnm{Cressie},~\bfnm{Noel A.~C.}\binits{N.~A.~C.}}
(\byear{1993}).
\btitle{Statistics for Spatial Data}.
\bseries{Wiley Series in Probability and Mathematical Statistics: Applied
Probability and Statistics}.
\bpublisher{Wiley}, \blocation{New York}.
\bid{mr={1239641}}
\bptok{imsref}%
\end{bbook}
\endbibitem

\bibitem[\protect\citeauthoryear{Dawid}{1981}]{Daw81}
\begin{barticle}[mr]
\bauthor{\bsnm{Dawid},~\bfnm{A.~P.}\binits{A.~P.}}
(\byear{1981}).
\btitle{Some matrix-variate distribution theory: Notational considerations and
a~{B}ayesian application}.
\bjournal{Biometrika}
\bvolume{68}
\bpages{265--274}.
\bid{doi={10.1093/biomet/68.1.265}, issn={0006-3444}, mr={0614963}}
\bptok{imsref}%
\end{barticle}
\endbibitem

\bibitem[\protect\citeauthoryear{Debarsy, Ertur and LeSage}{2012}]{DebErtLeS12}
\begin{barticle}[mr]
\bauthor{\bsnm{Debarsy},~\bfnm{Nicolas}\binits{N.}},
\bauthor{\bsnm{Ertur},~\bfnm{Cem}\binits{C.}} \AND
\bauthor{\bsnm{LeSage},~\bfnm{James~P.}\binits{J.~P.}}
(\byear{2012}).
\btitle{Interpreting dynamic space--time panel data models}.
\bjournal{Stat. Methodol.}
\bvolume{9}
\bpages{158--171}.
\bid{doi={10.1016/j.stamet.2011.02.002}, issn={1572-3127}, mr={2863605}}
\bptok{imsref}%
\end{barticle}
\endbibitem

\bibitem[\protect\citeauthoryear{Di~Giacinto et~al.}{2005}]{DiGetal05}
\begin{barticle}[mr]
\bauthor{\bsnm{Di~Giacinto},~\bfnm{Valter}\binits{V.}},
\bauthor{\bsnm{Dryden},~\bfnm{Ian}\binits{I.}},
\bauthor{\bsnm{Ippoliti},~\bfnm{Luigi}\binits{L.}} \AND
\bauthor{\bsnm{Romagnoli},~\bfnm{Luca}\binits{L.}}
(\byear{2005}).
\btitle{Linear smoothing of noisy spatial temporal series}.
\bjournal{J. Math. Stat.}
\bvolume{1}
\bpages{299--311}.
\bid{issn={1549-3644}, mr={2400983}}
\bptok{imsref}%
\end{barticle}
\endbibitem

\bibitem[\protect\citeauthoryear{Durbin and Koopman}{2001}]{DurKoo01}
\begin{bbook}[mr]
\bauthor{\bsnm{Durbin},~\bfnm{J.}\binits{J.}} \AND
\bauthor{\bsnm{Koopman},~\bfnm{S.~J.}\binits{S.~J.}}
(\byear{2001}).
\btitle{Time Series Analysis by State Space Methods}.
\bseries{Oxford Statistical Science Series}
\bvolume{24}.
\bpublisher{Oxford Univ. Press}, \blocation{Oxford}.
\bid{mr={1856951}}
\bptok{imsref}%
\end{bbook}
\endbibitem

\bibitem[\protect\citeauthoryear{Elhorst}{2001}]{Elh01}
\begin{barticle}[auto:STB|2013/02/26|09:05:15]
\bauthor{\bsnm{Elhorst},~\bfnm{J.~P.}\binits{J.~P.}}
(\byear{2001}).
\btitle{Dynamic models in space and time}.
\bjournal{Geographical Analysis}
\bvolume{33}
\bpages{119--140}.
\bptok{imsref}%
\end{barticle}
\endbibitem

\bibitem[\protect\citeauthoryear{Engle, Hendry and Richard}{1983}]{EngHenRic83}
\begin{barticle}[mr]
\bauthor{\bsnm{Engle},~\bfnm{Robert~F.}\binits{R.~F.}},
\bauthor{\bsnm{Hendry},~\bfnm{David~F.}\binits{D.~F.}} \AND
\bauthor{\bsnm{Richard},~\bfnm{Jean-Fran{\c{c}}ois}\binits{J.-F.}}
(\byear{1983}).
\btitle{Exogeneity}.
\bjournal{Econometrica}
\bvolume{51}
\bpages{277--304}.
\bid{doi={10.2307/1911990}, issn={0012-9682}, mr={0688727}}
\bptok{imsref}%
\end{barticle}
\endbibitem

\bibitem[\protect\citeauthoryear{ESRI}{2009}]{ESRI09}
\begin{bmisc}[auto:STB|2013/02/26|09:05:15]
\borganization{ESRI}.
(\byear{2009}).
\bhowpublished{ArcMap 9.2. Environmental Systems Resource Institute, Redlands,
California}.
\bptok{imsref}%
\end{bmisc}
\endbibitem

\bibitem[\protect\citeauthoryear{Fr{\"u}hwirth-Schnatter}{1994}]{Fru94}
\begin{barticle}[mr]
\bauthor{\bsnm{Fr{\"u}hwirth-Schnatter},~\bfnm{Sylvia}\binits{S.}}
(\byear{1994}).
\btitle{Data augmentation and dynamic linear models}.
\bjournal{J.~Time Series Anal.}
\bvolume{15}
\bpages{183--202}.
\bid{doi={10.1111/j.1467-9892.1994.tb00184.x}, issn={0143-9782}, mr={1263889}}
\bptok{imsref}%
\end{barticle}
\endbibitem

\bibitem[\protect\citeauthoryear{Gallin}{2008}]{Gal08}
\begin{barticle}[auto:STB|2013/02/26|09:05:15]
\bauthor{\bsnm{Gallin},~\bfnm{J.}\binits{J.}}
(\byear{2008}).
\btitle{The long run relationship between housing prices and income: Evidence
from local housing markets}.
\bjournal{Real Estate Economics}
\bvolume{36}
\bpages{635--658}.
\bptok{imsref}%
\end{barticle}
\endbibitem

\bibitem[\protect\citeauthoryear{Gelfand and Ghosh}{1998}]{GelGho98}
\begin{barticle}[mr]
\bauthor{\bsnm{Gelfand},~\bfnm{Alan~E.}\binits{A.~E.}} \AND
\bauthor{\bsnm{Ghosh},~\bfnm{Sujit~K.}\binits{S.~K.}}
(\byear{1998}).
\btitle{Model choice: A minimum posterior predictive loss approach}.
\bjournal{Biometrika}
\bvolume{85}
\bpages{1--11}.
\bid{doi={10.1093/biomet/85.1.1}, issn={0006-3444}, mr={1627258}}
\bptok{imsref}%
\end{barticle}
\endbibitem

\bibitem[\protect\citeauthoryear{Gelman}{1996}]{Gel96}
\begin{bmisc}[auto:STB|2013/02/26|09:05:15]
\bauthor{\bsnm{Gelman},~\bfnm{A.}\binits{A.}}
(\byear{1996}).
\bhowpublished{Inference and Monitoring Convergence. In \textit{Introducing Markov Chain Monte Carlo}.}
\bptok{imsref}%
\end{bmisc}
\endbibitem

\bibitem[\protect\citeauthoryear{George, Sun and Ni}{2008}]{GeoSunNi08}
\begin{barticle}[mr]
\bauthor{\bsnm{George},~\bfnm{Edward~I.}\binits{E.~I.}},
\bauthor{\bsnm{Sun},~\bfnm{Dongchu}\binits{D.}} \AND
\bauthor{\bsnm{Ni},~\bfnm{Shawn}\binits{S.}}
(\byear{2008}).
\btitle{Bayesian stochastic search for {VAR} model restrictions}.
\bjournal{J. Econometrics}
\bvolume{142}
\bpages{553--580}.
\bid{doi={10.1016/j.jeconom.2007.08.017}, issn={0304-4076}, mr={2408749}}
\bptok{imsref}%
\end{barticle}
\endbibitem


\bibitem[\protect\citeauthoryear{Geweke}{1992}]{Gew92}
\begin{bincollection}[mr]
\bauthor{\bsnm{Geweke},~\bfnm{John}\binits{J.}}
(\byear{1992}).
\btitle{Evaluating the accuracy of sampling-based approaches to the calculation
of posterior moments}.
In \bbooktitle{Bayesian Statistics, 4 ({P}e\~n\'\i Scola, 1991)}
(\beditor{\binits{J.}\bfnm{J.} \bsnm{Bernardo}},
\beditor{\binits{J.}\bfnm{J.} \bsnm{Berger}},
\beditor{\binits{A.}\bfnm{A.} \bsnm{Dawid}}
\AND
\beditor{\binits{A.}\bfnm{A.} \bsnm{Smith}}, eds.)
\bpages{169--193}.
\bpublisher{Oxford Univ. Press}, \blocation{New York}.
\bid{mr={1380276}}
\bptok{imsref}%
\end{bincollection}
\endbibitem

\bibitem[\protect\citeauthoryear{Gilks, Richardson and
Spiegelhalter}{1996}]{GilRicSpi96}
\begin{bbook}[mr]
\beditor{\bsnm{Gilks},~\bfnm{W.~R.}\binits{W.~R.}},
\beditor{\bsnm{Richardson},~\bfnm{S.}\binits{S.}} \AND
\beditor{\bsnm{Spiegelhalter},~\bfnm{D.~J.}\binits{D.~J.}}, eds.
(\byear{1996}).
\btitle{Markov Chain {M}onte {C}arlo in Practice}.
\bseries{Interdisciplinary Statistics}.
\bpublisher{Chapman \& Hall}, \blocation{London}.
\bid{mr={1397966}}
\bptok{imsref}%
\end{bbook}
\endbibitem

\bibitem[\protect\citeauthoryear{Giussani and Hadjimatheou}{1991}]{GiuHad91}
\begin{barticle}[auto:STB|2013/02/26|09:05:15]
\bauthor{\bsnm{Giussani},~\bfnm{B.}\binits{B.}} \AND
\bauthor{\bsnm{Hadjimatheou},~\bfnm{G.}\binits{G.}}
(\byear{1991}).
\btitle{Modeling regional housing prices in the United Kingdom}.
\bjournal{Papers In Regional Science}
\bvolume{70}
\bpages{201--219}.
\bptok{imsref}%
\end{barticle}
\endbibitem

\bibitem[\protect\citeauthoryear{Gourieroux and Monfort}{1997}]{GouMon97}
\begin{bbook}[auto:STB|2013/02/26|09:05:15]
\bauthor{\bsnm{Gourieroux},~\bfnm{C.~S.}\binits{C.~S.}} \AND
\bauthor{\bsnm{Monfort},~\bfnm{A.}\binits{A.}}
(\byear{1997}).
\btitle{Time Series and Dynamic Models}.
\bpublisher{Cambridge Univ. Press}, \blocation{Cambridge}.
\bptok{imsref}%
\end{bbook}
\endbibitem

\bibitem[\protect\citeauthoryear{Holly, Pesaran and
Yamagata}{2010}]{HolPesYam10}
\begin{barticle}[mr]
\bauthor{\bsnm{Holly},~\bfnm{Sean}\binits{S.}},
\bauthor{\bsnm{Pesaran},~\bfnm{M.~Hashem}\binits{M.~H.}} \AND
\bauthor{\bsnm{Yamagata},~\bfnm{Takashi}\binits{T.}}
(\byear{2010}).
\btitle{A spatio-temporal model of house prices in the {USA}}.
\bjournal{J. Econometrics}
\bvolume{158}
\bpages{160--173}.
\bid{doi={10.1016/j.jeconom.2010.03.040}, issn={0304-4076}, mr={2671272}}
\bptok{imsref}%
\end{barticle}
\endbibitem

\bibitem[\protect\citeauthoryear{Ippoliti, Valentini and
Gamerman}{2012}]{IppValGam12}
\begin{barticle}[mr]
\bauthor{\bsnm{Ippoliti},~\bfnm{L.}\binits{L.}},
\bauthor{\bsnm{Valentini},~\bfnm{P.}\binits{P.}} \AND
\bauthor{\bsnm{Gamerman},~\bfnm{D.}\binits{D.}}
(\byear{2012}).
\btitle{Space-time modelling of coupled spatio-temporal environmental
variables}.
\bjournal{J. R. Stat. Soc. Ser. C. Appl. Stat.}
\bvolume{61}
\bpages{175--200}.
\bid{doi={10.1111/j.1467-9876.2011.01011.x}, issn={0035-9254}, mr={2905058}}
\bptok{imsref}%
\end{barticle}
\endbibitem

\bibitem[\protect\citeauthoryear{Jochmann et~al.}{2013}]{Jocetal11}
\begin{barticle}[auto:STB|2013/02/26|09:05:15]
\bauthor{\bsnm{Jochmann},~\bfnm{M.}\binits{M.}},
\bauthor{\bsnm{Koop},~\bfnm{G.}\binits{G.}},
\bauthor{\bsnm{Leon-Gonzalez},~\bfnm{R.}\binits{R.}} \AND
\bauthor{\bsnm{Strachan},~\bfnm{R.}\binits{R.}}
(\byear{2013}).
\btitle{Stochastic search variable selection in vector error correction models
with an application to a model of the UK macroeconomy}.
\bjournal{J. Appl. Econometrics}
\bvolume{28}
\bpages{62--81}.
\bid{doi={10.1002/jae.1238}}
\bptok{imsref}%
\end{barticle}
\endbibitem

\bibitem[\protect\citeauthoryear{Johansen}{1988}]{Joh88}
\begin{barticle}[mr]
\bauthor{\bsnm{Johansen},~\bfnm{S{\o}ren}\binits{S.}}
(\byear{1988}).
\btitle{Statistical analysis of cointegration vectors}.
\bjournal{J. Econom. Dynam. Control}
\bvolume{12}
\bpages{231--254}.
\bid{doi={10.1016/0165-1889(88)90041-3}, issn={0165-1889}, mr={0986516}}
\bptok{imsref}%
\end{barticle}
\endbibitem

\bibitem[\protect\citeauthoryear{Jones et~al.}{2006}]{Jonetal06}
\begin{barticle}[mr]
\bauthor{\bsnm{Jones},~\bfnm{Galin~L.}\binits{G.~L.}},
\bauthor{\bsnm{Haran},~\bfnm{Murali}\binits{M.}},
\bauthor{\bsnm{Caffo},~\bfnm{Brian~S.}\binits{B.~S.}} \AND
\bauthor{\bsnm{Neath},~\bfnm{Ronald}\binits{R.}}
(\byear{2006}).
\btitle{Fixed-width output analysis for {M}arkov chain {M}onte {C}arlo}.
\bjournal{J. Amer. Statist. Assoc.}
\bvolume{101}
\bpages{1537--1547}.
\bid{doi={10.1198/016214506000000492}, issn={0162-1459}, mr={2279478}}
\bptok{imsref}%
\end{barticle}
\endbibitem

\bibitem[\protect\citeauthoryear{Kass and Raftery}{1995}]{KasRaf95}
\begin{barticle}[auto:STB|2013/02/26|09:05:15]
\bauthor{\bsnm{Kass},~\bfnm{R.~E.}\binits{R.~E.}} \AND
\bauthor{\bsnm{Raftery},~\bfnm{A.~E.}\binits{A.~E.}}
(\byear{1995}).
\btitle{Bayes factors}.
\bjournal{J. Amer. Statist. Assoc.}
\bvolume{90}
\bpages{773--795}.
\bptok{imsref}%
\end{barticle}
\endbibitem

\bibitem[\protect\citeauthoryear{Kim, Sun and Tsutakawa}{2001}]{KimSunTsu01}
\begin{barticle}[mr]
\bauthor{\bsnm{Kim},~\bfnm{Hoon}\binits{H.}},
\bauthor{\bsnm{Sun},~\bfnm{Dongchu}\binits{D.}} \AND
\bauthor{\bsnm{Tsutakawa},~\bfnm{Robert~K.}\binits{R.~K.}}
(\byear{2001}).
\btitle{A bivariate {B}ayes method for improving the estimates of mortality
rates with a twofold conditional autoregressive model}.
\bjournal{J.~Amer. Statist. Assoc.}
\bvolume{96}
\bpages{1506--1521}.
\bid{doi={10.1198/016214501753382408}, issn={0162-1459}, mr={1946594}}
\bptok{imsref}%
\end{barticle}
\endbibitem

\bibitem[\protect\citeauthoryear{Koop, Le{\'o}n-Gonz{\'a}lez and
Strachan}{2010}]{KooLeoStr10}
\begin{barticle}[mr]
\bauthor{\bsnm{Koop},~\bfnm{Gary}\binits{G.}},
\bauthor{\bsnm{Le{\'o}n-Gonz{\'a}lez},~\bfnm{Roberto}\binits{R.}} \AND
\bauthor{\bsnm{Strachan},~\bfnm{Rodney~W.}\binits{R.~W.}}
(\byear{2010}).
\btitle{Efficient posterior simulation for cointegrated models with priors on
the cointegration space}.
\bjournal{Econometric Rev.}
\bvolume{29}
\bpages{224--242}.
\bid{doi={10.1080/07474930903382208}, issn={0747-4938}, mr={2747499}}
\bptnote{check year}%
\bptok{imsref}%
\end{barticle}
\endbibitem

\bibitem[\protect\citeauthoryear{Koop et~al.}{2006}]{Kooetal06}
\begin{bincollection}[auto:STB|2013/02/26|09:05:15]
\bauthor{\bsnm{Koop},~\bfnm{G.~M.}\binits{G.~M.}},
\bauthor{\bsnm{Strachan},~\bfnm{R.~W.}\binits{R.~W.}},
\bauthor{\bsnm{Van~Dijk},~\bfnm{H.}\binits{H.}} \AND
\bauthor{\bsnm{Villani},~\bfnm{M.}\binits{M.}}
(\byear{2006}).
\btitle{Bayesian approaches to cointegration}.
In \bbooktitle{The Palgrave Handbook of Theoretical Econometrics}
\bpages{871--898}.
\bpublisher{Palgrave Macmillan}, \blocation{Basingstoke, UK}.
\bptok{imsref}%
\end{bincollection}
\endbibitem

\bibitem[\protect\citeauthoryear{Kuethe and Pede}{2011}]{KuePed11}
\begin{barticle}[auto:STB|2013/02/26|09:05:15]
\bauthor{\bsnm{Kuethe},~\bfnm{T.}\binits{T.}} \AND
\bauthor{\bsnm{Pede},~\bfnm{V.}\binits{V.}}
(\byear{2011}).
\btitle{Regional housing price cycles: A spatio-temporal analysis using US
state-level data}.
\bjournal{Regional Studies}
\bvolume{45}
\bpages{563--574}.
\bptok{imsref}%
\end{barticle}
\endbibitem

\bibitem[\protect\citeauthoryear{Lopes, Salazar and
Gamerman}{2008}]{LopSalGam08}
\begin{barticle}[mr]
\bauthor{\bsnm{Lopes},~\bfnm{Hedibert~Freitas}\binits{H.~F.}},
\bauthor{\bsnm{Salazar},~\bfnm{Esther}\binits{E.}} \AND
\bauthor{\bsnm{Gamerman},~\bfnm{Dani}\binits{D.}}
(\byear{2008}).
\btitle{Spatial dynamic factor analysis}.
\bjournal{Bayesian Anal.}
\bvolume{3}
\bpages{759--792}.
\bid{doi={10.1214/08-BA329}, issn={1936-0975}, mr={2469799}}
\bptok{imsref}%
\end{barticle}
\endbibitem

\bibitem[\protect\citeauthoryear{Lopes and West}{2004}]{LopWes04}
\begin{barticle}[mr]
\bauthor{\bsnm{Lopes},~\bfnm{Hedibert~Freitas}\binits{H.~F.}} \AND
\bauthor{\bsnm{West},~\bfnm{Mike}\binits{M.}}
(\byear{2004}).
\btitle{Bayesian model assessment in factor analysis}.
\bjournal{Statist. Sinica}
\bvolume{14}
\bpages{41--67}.
\bid{issn={1017-0405}, mr={2036762}}
\bptok{imsref}%
\end{barticle}
\endbibitem

\bibitem[\protect\citeauthoryear{L{\"u}tkepohl}{2005}]{Lut05}
\begin{bbook}[mr]
\bauthor{\bsnm{L{\"u}tkepohl},~\bfnm{Helmut}\binits{H.}}
(\byear{2005}).
\btitle{New Introduction to Multiple Time Series Analysis}.
\bpublisher{Springer}, \blocation{Berlin}.
\bid{mr={2172368}}
\bptok{imsref}%
\end{bbook}
\endbibitem

\bibitem[\protect\citeauthoryear{Malpezzi}{1999}]{Mal99}
\begin{barticle}[auto:STB|2013/02/26|09:05:15]
\bauthor{\bsnm{Malpezzi},~\bfnm{S.}\binits{S.}}
(\byear{1999}).
\btitle{A simple error correction model of housing prices}.
\bjournal{Journal of Housing Economics}
\bvolume{8}
\bpages{27--62}.
\bptok{imsref}%
\end{barticle}
\endbibitem

\bibitem[\protect\citeauthoryear{Mardia}{1988}]{Mar88}
\begin{barticle}[mr]
\bauthor{\bsnm{Mardia},~\bfnm{K.~V.}\binits{K.~V.}}
(\byear{1988}).
\btitle{Multidimensional multivariate {G}aussian {M}arkov random fields with
application to image processing}.
\bjournal{J. Multivariate Anal.}
\bvolume{24}
\bpages{265--284}.
\bid{doi={10.1016/0047-259X(88)90040-1}, issn={0047-259X}, mr={0926357}}
\bptok{imsref}%
\end{barticle}
\endbibitem

\bibitem[\protect\citeauthoryear{Mardia, Kent and Bibby}{1979}]{MarKenBib79}
\begin{bbook}[mr]
\bauthor{\bsnm{Mardia},~\bfnm{Kantilal~Varichand}\binits{K.~V.}},
\bauthor{\bsnm{Kent},~\bfnm{John~T.}\binits{J.~T.}} \AND
\bauthor{\bsnm{Bibby},~\bfnm{John~M.}\binits{J.~M.}}
(\byear{1979}).
\btitle{Multivariate Analysis}.
\bpublisher{Academic Press},
\blocation{London}.
\bid{mr={0560319}}
\bptok{imsref}%
\end{bbook}
\endbibitem

\bibitem[\protect\citeauthoryear{Meen}{1999}]{Mee99}
\begin{barticle}[auto:STB|2013/02/26|09:05:15]
\bauthor{\bsnm{Meen},~\bfnm{G.}\binits{G.}}
(\byear{1999}).
\btitle{Regional house prices and the ripple effect: A new interpretation}.
\bjournal{Housing Studies}
\bvolume{14}
\bpages{733--753}.
\bptok{imsref}%
\end{barticle}
\endbibitem

\bibitem[\protect\citeauthoryear{Meen}{2001}]{Mee01}
\begin{bbook}[auto:STB|2013/02/26|09:05:15]
\bauthor{\bsnm{Meen},~\bfnm{G.}\binits{G.}}
(\byear{2001}).
\btitle{Modelling Spatial Housing Markets: Theory, Analysis and Policy}.
\bpublisher{Kluwer}, \blocation{Dordrecht, The Netherlands}.
\bptok{imsref}%
\end{bbook}
\endbibitem

\bibitem[\protect\citeauthoryear{Moench and Ng}{2011}]{MoeNg11}
\begin{barticle}[auto:STB|2013/02/26|09:05:15]
\bauthor{\bsnm{Moench},~\bfnm{E.}\binits{E.}} \AND
\bauthor{\bsnm{Ng},~\bfnm{S.}\binits{S.}}
(\byear{2011}).
\btitle{A hierarchical factor analysis of U.S. housing market dynamics}.
\bjournal{Econom. J.}
\bvolume{14}
\bpages{C1--C24}.
\bptok{imsref}%
\end{barticle}
\endbibitem

\bibitem[\protect\citeauthoryear{Muellbauer and Murphy}{1997}]{MueMur97}
\begin{barticle}[auto:STB|2013/02/26|09:05:15]
\bauthor{\bsnm{Muellbauer},~\bfnm{J.}\binits{J.}} \AND
\bauthor{\bsnm{Murphy},~\bfnm{A.}\binits{A.}}
(\byear{1997}).
\btitle{Booms and busts in the UK housing market}.
\bjournal{Econom.~J.}
\bvolume{107}
\bpages{1701--1727}.
\bptok{imsref}%
\end{barticle}
\endbibitem

\bibitem[\protect\citeauthoryear{Osiewalski and Steel}{1996}]{OsiSte96}
\begin{barticle}[mr]
\bauthor{\bsnm{Osiewalski},~\bfnm{Jacek}\binits{J.}} \AND
\bauthor{\bsnm{Steel},~\bfnm{Mark F.~J.}\binits{M.~F.~J.}}
(\byear{1996}).
\btitle{A {B}ayesian analysis of exogeneity in models pooling time-series and
cross-sectional data}.
\bjournal{J. Statist. Plann. Inference}
\bvolume{50}
\bpages{187--206}.
\bid{doi={10.1016/0378-3758(95)00053-4}, issn={0378-3758}, mr={1396456}}
\bptok{imsref}%
\end{barticle}
\endbibitem

\bibitem[\protect\citeauthoryear{Pesaran}{2006}]{Pes06}
\begin{barticle}[mr]
\bauthor{\bsnm{Pesaran},~\bfnm{M.~Hashem}\binits{M.~H.}}
(\byear{2006}).
\btitle{Estimation and inference in large heterogeneous panels with a
multifactor error structure}.
\bjournal{Econometrica}
\bvolume{74}
\bpages{967--1012}.
\bid{doi={10.1111/j.1468-0262.2006.00692.x}, issn={0012-9682}, mr={2238209}}
\bptok{imsref}%
\end{barticle}
\endbibitem

\bibitem[\protect\citeauthoryear{Pfeifer and Deutsch}{1980}]{PfeDeu80}
\begin{barticle}[auto:STB|2013/02/26|09:05:15]
\bauthor{\bsnm{Pfeifer},~\bfnm{P.~E.}\binits{P.~E.}} \AND
\bauthor{\bsnm{Deutsch},~\bfnm{S.~J.}\binits{S.~J.}}
(\byear{1980}).
\btitle{A three-stage iterative procedure for space-time modeling}.
\bjournal{Technometrics}
\bvolume{22}
\bpages{35--47}.
\bptok{imsref}%
\end{barticle}
\endbibitem

\bibitem[\protect\citeauthoryear{Pfeifer and Deutsch}{1981}]{PfeDeu81}
\begin{barticle}[auto:STB|2013/02/26|09:05:15]
\bauthor{\bsnm{Pfeifer},~\bfnm{P.~E.}\binits{P.~E.}} \AND
\bauthor{\bsnm{Deutsch},~\bfnm{S.~J.}\binits{S.~J.}}
(\byear{1981}).
\btitle{Space-time ARMA modeling with contemporaneously correlated
innovations}.
\bjournal{Technometrics}
\bvolume{23}
\bpages{401--409}.
\bptok{imsref}%
\end{barticle}
\endbibitem

\bibitem[\protect\citeauthoryear{Primiceri}{2005}]{Pri05}
\begin{barticle}[mr]
\bauthor{\bsnm{Primiceri},~\bfnm{Giorgio~E.}\binits{G.~E.}}
(\byear{2005}).
\btitle{Time varying structural vector autoregressions and monetary policy}.
\bjournal{Rev. Econom. Stud.}
\bvolume{72}
\bpages{821--852}.
\bid{doi={10.1111/j.1467-937X.2005.00353.x}, issn={0034-6527}, mr={2148143}}
\bptok{imsref}%
\end{barticle}
\endbibitem

\bibitem[\protect\citeauthoryear{Rosenberg}{1973}]{Ros73}
\begin{barticle}[auto:STB|2013/02/26|09:05:15]
\bauthor{\bsnm{Rosenberg},~\bfnm{B.}\binits{B.}}
(\byear{1973}).
\btitle{Random coefficients models: The analysis of a cross-section of time
series by stochastically convergent parameter regression}.
\bjournal{Annals of Economic and Social Measurement}
\bvolume{60}
\bpages{399--428}.
\bptok{imsref}%
\end{barticle}
\endbibitem

\bibitem[\protect\citeauthoryear{Sain and Cressie}{2007}]{SaiCre07}
\begin{barticle}[mr]
\bauthor{\bsnm{Sain},~\bfnm{Stephan~R.}\binits{S.~R.}} \AND
\bauthor{\bsnm{Cressie},~\bfnm{Noel}\binits{N.}}
(\byear{2007}).
\btitle{A spatial model for multivariate lattice data}.
\bjournal{J.~Econometrics}
\bvolume{140}
\bpages{226--259}.
\bid{doi={10.1016/j.jeconom.2006.09.010}, issn={0304-4076}, mr={2395923}}
\bptok{imsref}%
\end{barticle}
\endbibitem

\bibitem[\protect\citeauthoryear{Sain, Furrer and Cressie}{2011}]{SaiFurCre11}
\begin{barticle}[mr]
\bauthor{\bsnm{Sain},~\bfnm{Stephan~R.}\binits{S.~R.}},
\bauthor{\bsnm{Furrer},~\bfnm{Reinhard}\binits{R.}} \AND
\bauthor{\bsnm{Cressie},~\bfnm{Noel}\binits{N.}}
(\byear{2011}).
\btitle{A spatial analysis of multivariate output from regional climate
models}.
\bjournal{Ann. Appl. Stat.}
\bvolume{5}
\bpages{150--175}.
\bid{doi={10.1214/10-AOAS369}, issn={1932-6157}, mr={2810393}}
\bptok{imsref}%
\end{barticle}
\endbibitem

\bibitem[\protect\citeauthoryear{Sims and Zha}{1999}]{SimZha99}
\begin{barticle}[mr]
\bauthor{\bsnm{Sims},~\bfnm{Christopher~A.}\binits{C.~A.}} \AND
\bauthor{\bsnm{Zha},~\bfnm{Tao}\binits{T.}}
(\byear{1999}).
\btitle{Error bands for impulse responses}.
\bjournal{Econometrica}
\bvolume{67}
\bpages{1113--1155}.
\bid{doi={10.1111/1468-0262.00071}, issn={0012-9682}, mr={1707461}}
\bptok{imsref}%
\end{barticle}
\endbibitem

\bibitem[\protect\citeauthoryear{Spiegelhalter et~al.}{2002}]{Spietal02}
\begin{barticle}[mr]
\bauthor{\bsnm{Spiegelhalter},~\bfnm{David~J.}\binits{D.~J.}},
\bauthor{\bsnm{Best},~\bfnm{Nicola~G.}\binits{N.~G.}},
\bauthor{\bsnm{Carlin},~\bfnm{Bradley~P.}\binits{B.~P.}} \AND
\bauthor{\bparticle{van~der} \bsnm{Linde},~\bfnm{Angelika}\binits{A.}}
(\byear{2002}).
\btitle{Bayesian measures of model complexity and fit}.
\bjournal{J. R. Stat. Soc. Ser. B Stat. Methodol.}
\bvolume{64}
\bpages{583--639}.
\bid{doi={10.1111/1467-9868.00353}, issn={1369-7412}, mr={1979380}}
\bptok{imsref}%
\end{barticle}
\endbibitem

\bibitem[\protect\citeauthoryear{Strickland et~al.}{2011}]{Stretal11}
\begin{barticle}[mr]
\bauthor{\bsnm{Strickland},~\bfnm{C.~M.}\binits{C.~M.}},
\bauthor{\bsnm{Simpson},~\bfnm{D.~P.}\binits{D.~P.}},
\bauthor{\bsnm{Turner},~\bfnm{I.~W.}\binits{I.~W.}},
\bauthor{\bsnm{Denham},~\bfnm{R.}\binits{R.}} \AND
\bauthor{\bsnm{Mengersen},~\bfnm{K.~L.}\binits{K.~L.}}
(\byear{2011}).
\btitle{Fast bayesian analysis of spatial
dynamic factor models for large space time data sets}.
\bjournal{J. R. Stat. Soc. Ser. C. Appl. Stat.}
\bvolume{60}
\bpages{1--16}.
\bptok{imsref}%
\end{barticle}
\endbibitem

\bibitem[\protect\citeauthoryear{Sugita}{2009}]{Sug09}
\begin{barticle}[auto:STB|2013/02/26|09:05:15]
\bauthor{\bsnm{Sugita},~\bfnm{K.}\binits{K.}}
(\byear{2009}).
\btitle{A Monte Carlo comparison of Bayesian testing for cointegration rank}.
\bjournal{Economics Bulletin}
\bvolume{29}
\bpages{2145--2151}.
\bptok{imsref}%
\end{barticle}
\endbibitem

\bibitem[\protect\citeauthoryear{van Dijk et~al.}{2011}]{vanetal11}
\begin{barticle}[auto:STB|2013/02/26|09:05:15]
\bauthor{\bparticle{van} \bsnm{Dijk},~\bfnm{B.}\binits{B.}},
\bauthor{\bsnm{Franses},~\bfnm{P.~H.}\binits{P.~H.}},
\bauthor{\bsnm{Paap},~\bfnm{R.}\binits{R.}} \AND \bauthor{\bparticle{van}
\bsnm{Dijk},~\bfnm{D.~J.~C.}\binits{D.~J.~C.}}
(\byear{2011}).
\btitle{Modeling regional house prices}.
\bjournal{Applied Economics}
\bvolume{43}
\bpages{2097--2110}.
\bptok{imsref}%
\end{barticle}
\endbibitem

\bibitem[\protect\citeauthoryear{Vermeulen and Van~Ommeren}{2009}]{VerVan09}
\begin{barticle}[auto:STB|2013/02/26|09:05:15]
\bauthor{\bsnm{Vermeulen},~\bfnm{W.}\binits{W.}} \AND
\bauthor{\bsnm{Van~Ommeren},~\bfnm{J.}\binits{J.}}
(\byear{2009}).
\btitle{Compensation of regional unemployment in housing markets}.
\bjournal{Economica}
\bvolume{76}
\bpages{71--88}.
\bptok{imsref}%
\end{barticle}
\endbibitem

\bibitem[\protect\citeauthoryear{Wang and Wall}{2003}]{WanWal03}
\begin{barticle}[pbm]
\bauthor{\bsnm{Wang},~\bfnm{Fujun}\binits{F.}} \AND
\bauthor{\bsnm{Wall},~\bfnm{Melanie~M.}\binits{M.~M.}}
(\byear{2003}).
\btitle{Generalized common spatial factor model}.
\bjournal{Biostatistics}
\bvolume{4}
\bpages{569--582}.
\bid{doi={10.1093/biostatistics/4.4.569}, issn={1465-4644}, pii={4/4/569},
pmid={14557112}}
\bptok{imsref}%
\end{barticle}
\endbibitem

\end{thebibliography}
\end{document}